\documentclass[12pt]{article}

\usepackage{graphicx}

\ifx\pdfoutput\undefined
    \relax
\else
    \usepackage{epstopdf}
\fi

\makeatletter
\def\numberbysection{\@addtoreset{equation}{section}
 	\def\theequation{\thesection.\arabic{equation}}}
\makeatother

\numberbysection

\newcommand{\be}{\begin{eqnarray}}
\newcommand{\ee}{\end{eqnarray}}
\newcommand{\non}{\nonumber}
\newcommand{\sgn}{\mathop{\rm sgn}\nolimits}

\begin{document}

\begin{titlepage}
\strut\hfill ITP Budapest Report No. 616\\
\strut\hfill UMTG--245
\vspace{.5in}
\begin{center}

\LARGE NLIE for hole excited states in the sine-Gordon model\\
\LARGE with two boundaries \\[1.0in]
\large Changrim Ahn\footnote{
       Department of Physics, Ewha Womans University, 
       Seoul 120-750, South Korea},
       Zolt\'an Bajnok\footnote{
       Theoretical Physics Research Group of the Hungarian Academy of Sciences, 
       E\"otv\"os University,
       H-1117 Budapest, P\'azm\'any P\'eter s\'et\'any 1/A, Hungary},
       Rafael I. Nepomechie\footnote{
       Physics Department, P.O. Box 248046, University of Miami,
       Coral Gables, FL 33124 USA},
       L\'aszl\'o Palla\footnote{
       Institute for Theoretical Physics, E\"otv\"os University,
       H-1117 Budapest, P\'azm\'any P\'eter s\'et\'any 1/A, Hungary}
   and G\'abor Tak\'acs${}^{2}$\\

\end{center}

\vspace{.5in}

\begin{abstract}
We derive a nonlinear integral equation (NLIE) for some bulk excited
states of the sine-Gordon model on a finite interval with general
integrable boundary interactions, including boundary terms
proportional to the first time derivative of the field.  We use this
NLIE to compute numerically the dimensions of these states as a
function of scale, and check the UV and IR limits analytically.  
We also find further support for the ground-state NLIE by comparison
with boundary conformal perturbation theory (BCPT), boundary
truncated conformal space approach (BTCSA) and the boundary analogue 
of the L\"uscher formula. 
\end{abstract}
\end{titlepage}

\setcounter{footnote}{0}

\section{Introduction}\label{sec:intro}

The nonlinear integral equation (NLIE) approach \cite{KBP, DDV1} is a
powerful tool for studying finite-size effects in the sine-Gordon
model with both periodic \cite{DDV1} - \cite{Fe} and Dirichlet
\cite{LMSS, ABR} boundary conditions.  A NLIE has recently been proposed
\cite{AN} for the ground state of the sine-Gordon model on a
finite interval with more general integrable boundary conditions
\cite{Sk1, GZ}, including new boundary terms proportional to the first
time derivative of the field ($\partial_{y}\varphi$).  We propose here
a NLIE for some bulk excited states of this model, using which we
numerically compute the dimensions of these states as a function of
scale (the product of the length of the interval and the soliton mass)
from ultraviolet (UV) to infrared (IR).  We perform checks of the UV
and IR limits analytically.  Other approaches to studying this model
(although without the $\partial_{y}\varphi$ boundary terms) have been
considered in \cite{BPT1}-\cite{LR}.

Our NLIE is based on the Bethe Ansatz solution \cite{Ne, CLSW, NR} of
the XXZ model with general (both diagonal and nondiagonal) boundary
terms \cite{dVGR}.  A significant limitation of this solution is that
the boundary parameters are not all independent, as they must satisfy
a linear constraint relating the left and right boundary parameters.
(Such a constraint does not arise in the case of diagonal boundary
terms \cite{Ga, ABBBQ, Sk2}.)  Consequently, our NLIE is applicable
only when the boundary parameters of the sine-Gordon model (including
the coefficients of the $\partial_{y}\varphi$ boundary terms) obey a
corresponding constraint.

Three different sets of boundary parameters are introduced in the
course of this paper: the UV parameters $(\mu_{\pm}\,,
\varphi_{0}^{\pm} \,, \kappa_{\pm})$ appearing in the boundary
sine-Gordon action; the IR parameters $(\eta_{\pm}\,,
\vartheta_{\pm}\,, \gamma_{\pm})$ appearing in the sine-Gordon boundary $S$
matrix; and the lattice parameters $(a_{\pm}\,, b_{\pm}\,, c_{\pm})$
appearing in the XXZ spin-chain Hamiltonian.
The relations between the continuum parameters $(\mu_{\pm}\,, \varphi_{0}^{\pm})$ 
and $(\eta_{\pm}\,, \vartheta_{\pm})$ are known \cite{BPT1, AZ1}.
An important challenge in our Bethe-Ansatz-based approach is to have
the correct relations between the lattice and continuum boundary
parameters.  Such relations were proposed in \cite{AN}.  The
consistency of the results presented here for the UV and IR limits of
excited states provides further support for those relations.

The outline of this article is as follows.  In Section \ref{sec:SG} we
collect some results about the sine-Gordon model on a finite interval
which we use later to compare with the NLIE results.  In particular,
we clarify various aspects of the $\partial_{y}\varphi$ boundary
terms: the periodicity of the coefficients $\kappa_{\pm}$
(\ref{kappaperiodicity}), and the dependence of the UV conformal
dimensions (\ref{DeltaCFT}) and of the boundary $S$ matrices (\ref{gammakappa}) 
on these parameters.  In
Section \ref{sec:lattice} we review the construction of the counting
function for the corresponding light-cone lattice model \cite{AN}, and
the corresponding expression for the Casimir energy (\ref{SGCasimir}).
Moreover, we derive the lattice counting equation
(\ref{latticecounteq}), which is valid also for the homogeneous
$(\Lambda =0)$ open XXZ spin chain.  In Section \ref{sec:NLIE} we
present the continuum NLIE (\ref{DDVeqn}) which follows from the
lattice counting function.  For simplicity, we restrict our attention
to source contributions from holes and special roots.  We also note
the relations (\ref{boundparamreltn1}), (\ref{cgamma}) between the
lattice and continuum boundary parameters, and the constraints
(\ref{continuumconstraints}), (\ref{odd}) that these parameters must
obey.  In Section \ref{subsec:UVNLIE} we analyze the UV limit.  We
give the NLIE result for the UV conformal dimensions of states with
arbitrary numbers of holes and special roots (\ref{DeltaNLIE}), and
show that it can be consistent with the CFT result (\ref{DeltaCFT})
for appropriate values of the boundary parameters.  In Section
\ref{subsec:IR} we analyze the IR limit.  In particular, we verify
that the IR limit of the NLIE for a one-hole state is equivalent to
the Yang equation for a particle on an interval.  A noteworthy feature
of this computation is that the boundary $S$ matrices \cite{GZ} which
enter the Yang equation are not diagonal.  
Our numerical results, including comparisons with boundary conformal
perturbation theory (BCPT), boundary truncated conformal space
approach (BTCSA) \cite{YZ,DPTW} and the boundary L\"uscher 
formula  \cite{Lu,BPT3},  are presented in Section \ref{sec:numerical}.
Section \ref{sec:conclude} contains a brief summary and a list of some
remaining problems.  In Appendix A we present a discussion of BCPT and
BTCSA.

\section{The sine-Gordon model on a finite interval}\label{sec:SG}

In this Section, we collect some results about the sine-Gordon model
on a finite interval which will be needed later for making comparisons
with NLIE results.  In particular, we clarify various aspects of the
$\partial_{y}\varphi$ boundary terms: the periodicity of the
coefficients $\kappa_{\pm}$, the dependence of the UV conformal
dimensions on these parameters, and the dependence of the boundary $S$
matrices on these parameters.

\subsection{Action}

Following \cite{AN}, we consider the sine-Gordon quantum field theory
on the finite ``spatial'' interval $x \in \left[ x_{-} \,, x_{+}
\right]$, with Euclidean action
\be
{\cal A}_{E} = \int_{-\infty}^{\infty}dy 
\int_{x_{-}}^{x_{+}}dx\  A(\varphi \,, \partial_{\mu} \varphi) 
+ \int_{-\infty}^{\infty}dy \left[ 
B_{-}(\varphi \,, \partial_{y}\varphi )\Big\vert_{x=x_{-}} +
B_{+}(\varphi \,, \partial_{y}\varphi )\Big\vert_{x=x_{+}} \right] \,,
\label{SGaction}
\ee 
where the bulk terms are given by
\be 
A(\varphi \,, \partial_{\mu} \varphi) = 
{1\over 2}(\partial_{\mu} \varphi)^{2}
- \mu_{bulk} \cos (\beta \varphi) \,,
\label{SGbulkaction}
\ee 
and the boundary terms are given by \footnote{While in \cite{AN} 
the coefficients of $\partial_{y}\varphi$ are expressed in terms of the
parameters $\gamma_{\pm}$ in the boundary $S$ matrices
(\ref{matrixelements}), here we instead denote these coefficients by
new parameters $\kappa_{\pm}$.}
\be
B_{\pm}(\varphi \,, \partial_{y}\varphi ) = 
-\mu_{\pm} \cos( {\beta\over 2} (\varphi - \varphi_{0}^{\pm}))
\pm i \kappa_{\pm} \partial_{y}\varphi  \,.
\label{SGboundaction}
\ee
As noted in \cite{AN}, this action is similar to the one considered by
Ghoshal and Zamolodchikov \cite{GZ}, except that now there are two
boundaries instead of one, and the boundary action
(\ref{SGboundaction}) contains an additional term depending on the
``time'' derivative of the field.  In the one-boundary case, such a
term can be eliminated by adding to the bulk action
(\ref{SGbulkaction}) a term proportional to $\partial_{x} \partial_{y}
\varphi$, which has no effect on the classical equations of motion.
However, in the two-boundary case, one can eliminate in this way only
one of the two $\kappa_{\pm}$ parameters (say, $\kappa_{+}$), which
results in a shift of the other ($\kappa_{-} \mapsto \kappa_{-} -
\kappa_{+}$). 

The $\kappa_{\pm}$ parameters are real.  The factor of $i$ in the
$\partial_{y}\varphi$ terms in (\ref{SGboundaction}) (which was missed
in \cite{AN}) is introduced by the Wick rotation from Minkowski to
Euclidean space.  Indeed, the Minkowski-space action is given by
${\cal A}_{M} = \int_{-\infty}^{\infty}dt\ L_{M}$, with
\be
L_{M} = \int_{x_{-}}^{x_{+}}dx\ 
{1\over 2}\left(
(\partial_{t}\varphi)^{2} - (\partial_{x}\varphi)^{2}\right)
- \kappa_{+} \partial_{t} \varphi(x_{+}) 
+ \kappa_{-} \partial_{t} \varphi(x_{-}) 
+ \ldots \,,
\label{Minkaction}
\ee
where the ellipsis ($\cdots$) represents the mass terms (proportional
to $\mu_{bulk}$ or $\mu_{\pm}$) which we have suppressed for brevity.
With $\varphi$ and $\kappa_{\pm}$ real, the Minkowski-space 
action is real, as is necessary. Rotating to Euclidean space $t=-i y$, 
$\partial_{t} \varphi = i \partial_{y}\varphi$, we see that 
\be
L_{M} = -\left\{ 
\int_{x_{-}}^{x_{+}}dx\ 
{1\over 2}\left( 
(\partial_{y}\varphi)^{2} + (\partial_{x}\varphi)^{2}\right)
+ i\kappa_{+} \partial_{y}\varphi(x_{+}) 
-i \kappa_{-} \partial_{y}\varphi(x_{-}) 
+ \ldots \right\} \equiv -L_{E}\,.
\ee
Hence, ${\cal A}_{M} = \int_{-\infty}^{\infty}(-i dy)\ (-L_{E}) =
i {\cal A}_{E}$, with the Euclidean action 
${\cal A}_{E} = \int_{-\infty}^{\infty}dy\ L_{E}$
given by Eqs. (\ref{SGaction})-(\ref{SGboundaction}). As usual, 
$e^{i {\cal A}_{M}} = e^{-{\cal A}_{E}}$.

An important observation is that the $\kappa_{\pm}$
parameters are periodic, with periodicity $\beta/2$.  Indeed, 
first observe that the action (\ref{SGaction})-(\ref{SGboundaction}) 
has the periodicity \footnote{Although the bulk terms (\ref{SGbulkaction}) 
have the periodicity $\varphi(x,y) \mapsto \varphi(x,y) + {2\pi\over 
\beta}$, the boundary terms (\ref{SGboundaction}) have only the 
reduced periodicity (\ref{fieldperiodicity}).}
\be
\varphi(x,y) \mapsto \varphi(x,y) + {4\pi\over \beta} \,.
\label{fieldperiodicity}
\ee
The contribution from the $\partial_{y}\varphi$ boundary terms
to $e^{-{\cal A}_{E}}$ in the Euclidean path integral 
$\int {\cal D}\varphi\ e^{-{\cal A}_{E}}$
is evidently given by
\be
e^{-i \left( \kappa_{+} \Delta \varphi(x_{+}) - \kappa_{-} \Delta 
\varphi(x_{-}) \right)} \,,
\label{phase1}
\ee
where
\be
\Delta \varphi(x) \equiv 
\int_{-\infty}^{\infty}dy\ \partial_{y}\varphi(x,y)
= \varphi(x,y=\infty) - \varphi(x,y=-\infty) \,.
\ee
Let us compactify the $y$ axis to a circle, so that $y=-\infty$ and
$y=\infty$ correspond to the same point.  It follows that the
sine-Gordon field on the boundary at $y=-\infty$ must be identified
with that at $y=\infty$, up to the periodicity (\ref{fieldperiodicity}).
Hence,
\be
\Delta \varphi(x_{\pm}) = {4\pi\over \beta} n_{\pm} \,,
\ee
where $n_{\pm}$ are integers. It follows that the contribution 
(\ref{phase1}) to $e^{-{\cal A}_{E}}$ becomes
\be
e^{-{4\pi i\over \beta} \left( \kappa_{+} n_{+} - \kappa_{-} n_{-}\right)} 
\,,
\label{phase2}
\ee
which has the periodicity
\be
\kappa_{\pm} \mapsto \kappa_{\pm} + {\beta\over 2} \,.
\label{kappaperiodicity}
\ee

We recall here that it is useful to introduce the parameters $\lambda$
and $\nu$ which are related to the bulk coupling constant $\beta$,
\be
\lambda = {8\pi\over \beta^{2}} - 1 = {1\over \nu -1} \,.
\label{bulkparamreltn}
\ee
Hence, the attractive ($0 < \beta^{2} < 4 \pi$) and repulsive ($4 \pi
< \beta^{2} < 8 \pi$) regimes correspond to the ranges $1 < \nu < 2$
and $\nu >2$, respectively.

\subsection{Ultraviolet limit}\label{subsec:UVCFT}

The sine-Gordon model (\ref{SGaction})-(\ref{SGboundaction}) can be
regarded as a perturbed $c=1$ boundary conformal field theory (CFT).
In the ultraviolet limit $\mu_{bulk}, \mu_{\pm} \rightarrow 0$, the
Minkowski-space Lagrangian is given by (see (\ref{Minkaction}))
\be
L_{M} = \int_{x_{-}}^{x_{+}}dx\ 
{1\over 2}\left(
(\partial_{t}\varphi)^{2} - (\partial_{x}\varphi)^{2}\right)
- \kappa_{+} \partial_{t} \varphi(x_{+}) 
+ \kappa_{-} \partial_{t} \varphi(x_{-}) \,.
\label{lagrangian}
\ee
It follows from the variational principle that $\varphi(x,t)$ obeys
the massless free field equation and Neumann boundary conditions,
\be
\left(\partial_{t}^{2}- \partial_{x}^{2}\right) \varphi(x,t) = 0 \,,
\qquad 
\partial_{x} \varphi(x,t) |_{x=x_{\pm}} = 0 \,.
\label{eom}
\ee 
Although the $\partial_{t} \varphi$ boundary terms do not affect
the central charge (they are ``marginal'' perturbations), they
modify the expression for the conformal dimension, which we now
proceed to compute by canonical quantization.

The canonical momentum $\Pi$ conjugate to $\varphi$ is given by
\be
\Pi = {\delta {\cal L}\over \delta (\partial_{t}\varphi)} =
\partial_{t}\varphi - \kappa_{+} \delta(x-x_{+}) 
+\kappa_{-} \delta(x-x_{-}) \,,
\label{momentum}
\ee
where ${\cal L}$ is the Lagrange density whose spatial integral
is the Lagrangian (\ref{lagrangian}). We expand $\varphi$ 
in terms of modes,
\be
\varphi(x,t) = \varphi_{0} + \pi_{0} {t\over L} 
+ {i\over \sqrt{\pi}} \sum_{n\ne 0} {1\over n} \alpha_{n} 
\cos (\pi n (x - x_{-})/L) e^{-i \pi n t/L} \,,
\label{expansion}
\ee
where $L= x_{+} - x_{-}$.  One can verify that this expression
satisfies the equations of motion (\ref{eom}).  The mode expansion for
$\Pi$ is obtained by substituting (\ref{expansion}) into
(\ref{momentum}).  Note that the momentum zero mode $\Pi_{0}$ is given
by
\be
\Pi_{0} = \int_{x_{-}}^{x_{+}}dx\  \Pi(x,t) = \pi_{0} 
 - \kappa_{+} + \kappa_{-} \,.
\label{momentumzeromode}
\ee 
The canonical equal-time commutation relations
\be
\left[ \Pi(x,t) \,, \varphi(x',t) \right] = -i \delta(x-x') \,, 
\qquad \left[ \Pi(x,t) \,, \Pi(x',t) \right] = 
\left[ \varphi(x,t) \,, \varphi(x',t) \right] = 0 \,,
\label{etcr}
\ee
imply that $\left[ \alpha_{n} \,, \alpha_{m} \right] = 
n \delta_{n+m,0}$ and
\be
\left[ \Pi_{0} \,, \varphi_{0} \right] = -i \,.
\ee

The Hamiltonian is given by
\be
H = \int_{x_{-}}^{x_{+}}dx\ {\cal H} \,, 
\qquad {\cal H} = \Pi\ \partial_{t}\varphi - {\cal L} \,.
\ee
Substituting the mode expansions, we obtain
\be
H = {1\over 2L} \pi_{0}^{2} + \mbox{ modes } \,,
\label{hamiltonian}
\ee
where ``modes'' represents the contribution of the 
oscillators $\alpha_{n}$. The wave functional of the zero mode is a 
plane wave,
\be
\Psi(\varphi_{0}) = e^{i \Pi_{0} \varphi_{0}} \,.
\ee

Let us now compactify the Boson on a circle with radius $r$, which
means that the theory is invariant under
\be
\varphi(x,t) \mapsto \varphi(x,t) + 2\pi r \,,
\ee
or equivalently, $\varphi_{0} \mapsto \varphi_{0} + 2\pi r$. 
Imposing this condition on the wave functional 
$\Psi(\varphi_{0}) = \Psi(\varphi_{0} + 2\pi r)$ implies
the quantization of the momentum zero mode,
\be
\Pi_{0} = {n\over r} \,,
\label{quantization}
\ee
where $n$ is an integer.  In view of (\ref{momentumzeromode}),
(\ref{hamiltonian}) and (\ref{quantization}), the zero-mode
contribution to the energy is
\be
E_{0\,, n} = {1\over 2L} \left(  {n\over r} + \kappa_{+} - \kappa_{-} 
\right)^{2} \,.
\ee
Comparing this result with the CFT result
\be
E_{0\,, n} = -{\pi\over 24L} (c_{eff}-1) = {\pi\over L}\Delta_{n} 
\ee
leads to the following expression for the conformal dimension
\be
\Delta_{n} = {1\over 2\pi} \left(  {n\over r} + \kappa_{+} - \kappa_{-} 
\right)^{2} \,.
\ee

For the boundary sine-Gordon model and its UV limit,  
the compactification radius must be 
\be
r = {2\over \beta} \,,
\ee
corresponding to the periodicity (\ref{fieldperiodicity}).
We conclude that $\Delta_{n}$ is given by
\be
\Delta_{n} &=& {1\over 2\pi} \left(  {n  \beta \over 2} + \kappa_{+} - \kappa_{-} 
\right)^{2} \non \\
&=& {1\over 4 \nu (\nu-1)} \left[  2n(\nu-1) 
+ {4\over \beta} (\nu-1)(\kappa_{+} - \kappa_{-}) 
\right]^{2} \,.
\label{DeltaCFT}
\ee
This result is consistent with the $\kappa_{\pm}$ periodicity
(\ref{kappaperiodicity}).  Also, this result is ``dual''
to the corresponding result for a free massless Boson with Dirichlet
boundary conditions \cite{Sa, AOS}.

\subsection{Boundary $S$ matrices}

Results from the theory on the left half line \cite{GZ}
imply that the right and left
boundary $S$ matrices $R(\theta\,; \eta_{\pm}, \vartheta_{\pm},
\gamma_{\pm})$ are given by 
\be
R(\theta\,; \eta, \vartheta, \gamma) = r_{0}(\theta)\ 
r_{1}(\theta\,; \eta, \vartheta)\ M(\theta\,; \eta, \vartheta, 
\gamma) \,,
\label{boundSmatrix}
\ee
where $M$ has matrix elements
\be
M(\theta\,; \eta, \vartheta, \gamma) =
\left( \begin{array}{cc}
m_{11} & m_{12} \\
m_{21} & m_{22}
	\end{array} \right) \,,
\ee
where 
\be
m_{11} &=& \cos \eta \cosh \vartheta \cosh (\lambda \theta)
+ i \sin \eta \sinh \vartheta  \sinh (\lambda \theta) \,, \non \\
m_{22} &=& \cos \eta \cosh \vartheta \cosh (\lambda \theta)
- i \sin \eta \sinh \vartheta  \sinh (\lambda \theta) \,, \non \\
m_{12} &=&i e^{i \gamma} \sinh(\lambda \theta) 
\cosh (\lambda \theta) \,, \non \\
m_{21} &=&i e^{-i \gamma} \sinh(\lambda \theta) 
\cosh (\lambda \theta) \,.
\label{matrixelements}
\ee 
Moreover, the scalar factors have the integral representations (see,
e.g., \cite{CSS2})
\be
r_{0}(\theta) &=& \exp \left\{ 2i\int_{0}^{\infty} {d\omega\over \omega}
\sin (2\theta \omega/ \pi) {\sinh ((\nu-2)\omega/2) \sinh(3\omega/2)\over
\sinh((\nu-1)\omega/2) \sinh(2\omega)} \right\} \,, \non \\
r_{1}(\theta\,; \eta, \vartheta) &=& {1\over \cos \eta \cosh \vartheta}
\sigma(\eta, \theta)\ \sigma(i\vartheta, \theta) \,,
\label{r0r1}
\ee
where
\be
\sigma(x, \theta) = \exp \left\{ 2\int_{0}^{\infty} {d\omega\over \omega}
\sin((i\pi -\theta) \omega/(2\pi)) \sin(\theta \omega/(2\pi))
{\cosh ((\nu-1)\omega x/\pi) \over
\sinh((\nu-1)\omega/2) \cosh(\omega/2)} \right\} \,.
\ee

Note the presence of the factors $e^{\pm i\gamma}$ in the off-diagonal
matrix elements $m_{12}$ and $m_{21}$, which are related to the
presence of the $\partial_{y}\varphi$ terms in the boundary action
(\ref{SGboundaction}), and which are absent in the case of a single
boundary \cite{GZ}.  In \cite{AN} an argument from \cite{GZ} was borrowed 
to determine the relation between the (real) parameters
$\gamma_{\pm}$ in the boundary $S$ matrix and the (real) parameters
$\kappa_{\pm}$ in the boundary action; namely (after correcting for
the missing $i$), $\gamma_{\pm}= \beta\kappa_{\pm}/ \pi$.  However,
this relation seems to be incorrect, since it would imply that the
factors $e^{i\gamma_{\pm}}$ do not have the periodicity
(\ref{kappaperiodicity}).  We propose here instead the
relation \footnote{The two relations coincide at the free 
Fermion point.}
\be
\gamma_{\pm}= {4 \pi \over \beta} \kappa_{\pm} \,,
\label{gammakappa}
\ee 
which implies that the factors $e^{i\gamma_{\pm}}$ (and
hence, the boundary $S$ matrix) do have the expected periodicity
(\ref{kappaperiodicity}). 

The relation (\ref{gammakappa}) for the right boundary can be
understood from elementary considerations. Indeed, for the
theory on the left half-line $x_{-} = -\infty \,, x_{+} =0$, the
Minkowski-space Lagrangian (\ref{lagrangian}) can be written in the
form
\be
L_{M} = L_{M}\left(\kappa_{+}=0\right)
-\kappa_{+}\partial_{t}\varphi(x=0) \,.
\label{minkL}
\ee
The amplitude for a process can be expressed using a path integral
of the form
\be 
\int \mathcal{D} \varphi\ \exp \left(i\int_{-\infty}^{\infty}dt\ 
L_{M}\right) 
= e^{-i\kappa_{+} \Delta \varphi}
\int \mathcal{D} \varphi\ \exp \left(i\int_{-\infty}^{\infty}dt\ 
L_{M} \left(\kappa_{+}=0\right)\right) \,,
\label{minkPI}
\ee
where now
\be
\Delta \varphi = \varphi(x=0,\, t=\infty)-\varphi(x=0,\, t=-\infty)
\,,
\ee 
with appropriate initial and final configurations of the field $\varphi$
at $t=\mp\infty$. For definiteness, we can fix the asymptotic condition
$\varphi(x=-\infty)=0$. For a process involving a soliton reflecting
back into a soliton, we have %(as part of the boundary conditions)
\be 
\varphi\left(x=0,t=\mp\infty\right)=\frac{2\pi}{\beta}
\qquad\Rightarrow\qquad \Delta \varphi=0 \,.
\ee 
Hence, for such processes the amplitude is independent of
$\kappa_{+}$; and this is also true for the reflection of an
antisoliton into an antisoliton.  For a soliton reflecting into an
antisoliton we have
\be 
\varphi\left(x=0,t=\mp\infty\right)=\pm\frac{2\pi}{\beta}
\qquad\Rightarrow\qquad \Delta \varphi=-\frac{4\pi}{\beta} \,,
\ee
which results in a phase factor
\be 
\exp \left(i\frac{4\pi}{\beta}\kappa_{+} \right) \,.
\ee
For an antisoliton reflecting into a soliton the resulting phase
factor is the inverse of the above. This leads to the relation
(\ref{gammakappa}) between the parameters in the Lagrangian and 
the reflection factor for the right boundary.

Similarly, for the left boundary, we consider the theory on the 
right half-line $x_{-} = 0 \,, x_{+} = \infty$. The corresponding
Lagrangian and path integral are given by (\ref{minkL}) and (\ref{minkPI})
with $\kappa_{+} \mapsto -\kappa_{-}$; and we now fix $\varphi(x=\infty)=0$. 
For a soliton reflecting into an antisoliton we have
\be 
\varphi\left(x=0,t=\mp\infty\right)=\mp\frac{2\pi}{\beta}
\qquad\Rightarrow\qquad \Delta \varphi=\frac{4\pi}{\beta} \,,
\ee
which results in a phase factor
\be 
\exp \left(i\frac{4\pi}{\beta}\kappa_{-} \right) \,,
\ee
and leads to the relation (\ref{gammakappa}) for the left boundary.

We find further support for relation (\ref{gammakappa})
from a study of the UV and IR limits of the NLIE in Sections 
\ref{subsec:UVNLIE} and \ref{subsec:IR}. 

The relation of the boundary $S$-matrix parameters $\eta_{\pm},
\vartheta_{\pm}$ to the parameters $\mu_{\pm} \,, \varphi_{0}^{\pm}$ 
in the boundary action (\ref{SGboundaction}) is given by \cite{BPT1, AZ1}
\be
\cos \left({\beta^{2}\over 8\pi}(\eta_{\pm} + i \vartheta_{\pm}) \right)
&=& {\mu_{\pm}\over \mu_{c}}  
e^{\mp {i\over 2}\beta \varphi_{0}^{\pm}} \,, \non \\
\cos \left({\beta^{2}\over 8\pi}(\eta_{\pm} - i \vartheta_{\pm}) \right)
&=& {\mu_{\pm}\over \mu_{c}} e^{\pm {i\over 2}\beta \varphi_{0}^{\pm}} \,,
\label{alyosha1}
\ee
where
\be
\mu_{c} = \sqrt{2 \mu_{bulk}\over
\sin \left({\beta^{2}\over 8}\right)} \,.
\label{alyosha2}
\ee 
Note that we have introduced an additional minus sign on one of the boundaries. 
That is, the UV-IR relation is different on the two boundaries, the
difference being in the sign of $\varphi_0$.
The two sets of boundary parameters $(\mu_{\pm}\,,
\varphi_{0}^{\pm} \,, \kappa_{\pm})$ and $(\eta_{\pm}\,,
\vartheta_{\pm}\,, \gamma_{\pm})$ can be regarded as ``UV'' and ``IR''
boundary parameters, respectively; hence, the relations
(\ref{gammakappa}), (\ref{alyosha1}), (\ref{alyosha2}) correspond to
UV-IR relations.

\section{The lattice counting function}\label{sec:lattice}

The light-cone lattice \cite{DDV3, DV, RS} version of the sine-Gordon
model is similar to the XXZ spin chain, the main difference being the
introduction of an alternating inhomogeneity parameter $\pm\Lambda$.
The solution \cite{Ne, CLSW} leads to the Bethe Ansatz equations
\cite{AN}
\be
h^{(+)}(\lambda_{j}) = I_{j}^{(+)} \,, \qquad  
j = 1 \,, \ldots \,, M^{(+)} \,,
\label{BAE}
\ee
where $\{ I_{j}^{(+)} \}$ are integers, and the lattice counting function
$h^{(+)}(\lambda)$ is given by \footnote{It should be clear from the context
whether the symbol $\lambda$ refers to the value (\ref{bulkparamreltn}) 
of the bulk coupling constant or to the rapidity variable, as in
(\ref{SGcounting}).}
\be
h^{(+)}(\lambda)&=&{1\over 2\pi}\Big\{ N \left[ q_{1}(\lambda + \Lambda)
+q_{1}(\lambda - \Lambda) \right]
+q_{1}(\lambda) + r_{1}(\lambda)
+ q_{2a_{-}-1}(\lambda) -  r_{1+2ib_{-}}(\lambda) \non \\
&+& q_{2a_{+}-1}(\lambda) -  r_{1+2ib_{+}}(\lambda)
- \sum_{k=1}^{M^{(+)}} \left[ q_{2}(\lambda - \lambda_{k}) +
q_{2}(\lambda + \lambda_{k})  \right] \Big\} \,.
\label{SGcounting}
\ee
The functions $q_{n}(\lambda)$ and $r_{n}(\lambda)$ are odd, and are
defined by \footnote{The branch cut of $\ln z$ is chosen along the
positive real axis; hence, $\ln (-1) = i \pi$.}
\be
q_{n}(\lambda) &=& \pi + i \ln 
{\sinh \left({\pi\over \nu} (\lambda + {i n\over 2}) \right) \over 
 \sinh \left({\pi\over \nu} (\lambda - {i n\over 2}) \right)} 
= 2 \tan^{-1}\left( \cot(n \pi/ (2\nu)) \tanh( \pi \lambda/\nu) \right)
\,, \non \\
r_{n}(\lambda) &=&  i \ln 
{\cosh \left({\pi\over \nu} (\lambda + {i n\over 2}) \right) \over 
 \cosh \left({\pi\over \nu} (\lambda - {i n\over 2}) \right)} \,.
\label{logfuncts}
\ee
The real lattice boundary parameters $a_{\pm}, b_{\pm}, c_{\pm}$
must satisfy the constraints
\be
a_{-} + a_{+} &=& \pm |c_{-} - c_{+}| + k \,, \non \\
b_{-} + b_{+} &=& 0 \,,
\label{latticeconstraints}
\ee
where the integer $k \in [-(N+1)\,, N+1]$ is even if $N$ is odd, and
is odd if $N$ is even.  The parameters $a_{\pm}$ can be restricted to
the fundamental domain $|2 a_{\pm}-1| < 2\nu$.  The number $M^{(+)}$ of
Bethe roots is given by
\be
M^{(+)}={1\over 2}(N-1+k) \,,
\label{Mvalue}
\ee
where $k$ is the integer in the constraint (\ref{latticeconstraints}).

The corresponding energy is given by \cite{LMSS, RS} \footnote{There 
is a misprint in the formula (2.29) in \cite{AN} for the boundary
energy of the (homogeneous) open XXZ chain: the first term in the
second line should be
\be
\sgn(2a_{\pm}-1){\sinh((\nu-|2a_{\pm}-1|)\omega/2) \over 
\sinh(\nu\omega/2)}  \,, \non
\ee
where $\sgn(n)$ is the function defined in (\ref{signfunc}).}
\be
E = - {1\over \delta} \sum_{j=1}^{M^{(+)}} 
\left[ a_{1}(\lambda_{j}+ \Lambda) 
+ a_{1}(\lambda_{j} - \Lambda) \right] \,,
\ee
where $\delta$ is the lattice spacing, and 
\be
a_n(\lambda) = {1\over 2\pi} {d \over d\lambda} q_n (\lambda)
= {1 \over \nu} 
{\sin (n \pi/\nu)\over \cosh(2 \pi \lambda/\nu) - \cos (n \pi/\nu)} \,.
% \,, \non \\
% b_n(\lambda) &=& {1\over 2\pi} {d \over d\lambda} r_n (\lambda)
% = -{1 \over \nu}  
% {\sin (n \pi/\nu)\over \cosh(2 \pi \lambda/\nu) + \cos (n \pi/\nu)} 
% \,. 
\ee 

For given values of the bulk and boundary parameters, the counting
function $h^{(+)}(\lambda)$ does not give all $2^{N}$ energy levels.
The remaining levels can be obtained from a corresponding counting
function $h^{(-)}(\lambda)$ with the boundary parameters negated,
\be
(a_{\pm}\,,  b_{\pm}) \mapsto (-a_{\pm}\,, -b_{\pm}) \,,
\label{negated}
\ee 
and with the number of Bethe roots equal to $M^{(-)} ={1\over
2}(N-1-k)$ \cite{NR}.  We shall refer to this other counting function
as the ``negated'' counting function.  Although below we generally
explicitly discuss only $h^{(+)}(\lambda)$, corresponding results hold
also for $h^{(-)}(\lambda)$.

Since the counting function (\ref{SGcounting}) is odd and has the
periodicity $\lambda \mapsto \lambda + i \nu$, we can restrict the
Bethe roots $\lambda_{j}$ to the following region of the complex
$\lambda$ plane \cite{ABR}
\be
\left\{ \Re e\  \lambda > 0 \,, \quad 
-{\nu\over 2} < \Im m\ \lambda \le {\nu\over 2} \right\} \bigcup 
\left\{ \Re e\  \lambda = 0 \,,  \quad 
0 < \Im m\ \lambda < {\nu\over 2} \right\} \,.
\label{region}
\ee
The origin ($\lambda=0$) is excluded since the corresponding
Bethe state would vanish. 

The summation over all the roots in the counting function involves the
function $q_{2}(\lambda)$, whose fundamental analyticity strip is
$|\Im m\ \lambda| < \min(1\,, \nu-1)$. Hence, it is useful
to classify the Bethe roots $\lambda_{j}$ 
in the region (\ref{region}) as either real, 
``close'' ($0 < |\Im m\ \lambda_{j}| < \min(1\,, \nu-1)$), 
or ``wide'' ($\min(1\,, \nu -1)  < |\Im m\ \lambda_{j}| < {\nu\over 2}$). 
Real solutions of $h^{(+)}(\lambda) = \mbox{ integer }$ 
which are not Bethe roots are called ``holes''. 
If an ``object'' (either a root or a hole) has rapidity $\lambda_{j}$ for
which the counting function is decreasing 
(${d\over d\lambda}h^{(+)}(\lambda_{j})< 0$), then the object is called 
``special''. We denote by $M_{R}$, $M_{C}$, $M_{W}$, $N_{H}$ and $N_{S}$ 
the number of real roots, close roots, wide roots, holes, and special 
objects, respectively. Note also that $h^{(+)}(\lambda)$ is continuous
on the real $\lambda$ axis. For further discussion about general 
properties of the counting function and the classification of roots 
and holes, see e.g. \cite{DDV2, ABR}.

We now proceed to derive a so-called lattice counting equation,
which relates $M_{C}$, $M_{W}$, $N_{H}$ and $N_{S}$ (but which is
independent of $M_{R}$ and $N$) for any Bethe state. To this end, we 
first compute the asymptotic limit of the counting function, and take 
its integer part,
\be
\lfloor h^{(+)}(\infty) \rfloor = M^{(+)} + 1 + {1\over 2}(s_{+}+s_{-}) - k 
+ \sgn(\nu-2) M_{W} + \lfloor {1\over 2} 
- {1\over \nu}\left( a_{+}+a_{-}-k \right) \rfloor
\,, \label{asymptotic}
\ee
where $\lfloor\quad \rfloor$ denotes integer part, and
$s_{\pm} = \sgn(a_{\pm} - {1\over 2})$, where
the sign function $\sgn(n)$ is defined as 
\be
\sgn(n) = \left\{ \begin{array}{c@{\quad : \quad} l}
{n\over |n|} & n \ne 0 \\
0 & n=0
\end{array} \right. \,.
\label{signfunc}
\ee
In obtaining the result (\ref{asymptotic}), we have used the facts
\be
q_{n}(\infty) &=& \sgn(n)\pi -  {n \pi\over \nu}\quad  
\mbox{ for }  0 < |n| < 2\nu \,,
\non \\
r_{n}(\infty) &=& - {n \pi\over \nu}\,,
\ee
as well as the relation (\ref{Mvalue}) to eliminate $N$, and the 
second constraint in (\ref{latticeconstraints}).  On the other
hand, one can argue that (see e.g. \cite{DDV2})
\be
\lfloor h^{(+)}(\infty) \rfloor = N_{H} + M_{R} - 2N_{S}
\,. \label{argue}
\ee 
Using the evident relation $M^{(+)} = M_{R} + M_{C} + M_{W}$ to eliminate 
$M_{R}$ on the RHS of (\ref{argue}), and then 
combining with (\ref{asymptotic}), we finally obtain the
lattice counting equation,
\be
N_{H} - 2N_{S} = M_{C} + 2M_{W} \mbox{step}(\nu-2) 
+1 +{1\over 2}(s_{+}+s_{-}) 
- k + \lfloor {1\over 2} 
- {1\over \nu}\left( a_{+}+a_{-} -k \right) \rfloor \,,
\label{latticecounteq}
\ee 
where the step function $\mbox{step}(n)$ is defined as
\be
\mbox{step}(n) = \left\{ \begin{array}{c@{\quad : \quad} l}
1 & n \ge 0 \\
0 & n<0
\end{array} \right. \,.
\ee
The lattice counting equation (\ref{latticecounteq}) is valid also for
the homogeneous $(\Lambda =0)$ open XXZ spin chain.

As a simple example of the utility of this result, consider (as
in \cite{AN}) the case that $N$ is even with $k=1$, and look for
purely real solutions with no holes or special roots $(M_{C} =
M_{W} = N_{H} = N_{S}= 0)$.  The lattice counting equation implies
\be
0 = {1\over 2}(s_{+}+s_{-})  + \lfloor {1\over 2} 
- {1\over \nu}\left( a_{+}+a_{-} -1 \right) \rfloor \,,
\ee 
which is a condition on the boundary parameters $a_{\pm}$ that is
necessary for such solutions to exist. Numerical checks suggest that 
this might also be a sufficient condition for the existence of such 
solutions.

\section{Nonlinear integral equation}\label{sec:NLIE}

The lattice NLIE can be derived from the lattice counting function
(\ref{SGcounting}) by standard manipulations \cite{DDV1} -
\cite{ABR}. The continuum limit consists of taking the number of
spins $N \rightarrow \infty$, the lattice spacing $\delta \rightarrow
0$, and the inhomogeneity parameter $\Lambda \rightarrow \infty$, in
such a way that the length $L = x_{+} - x_{-}$ and the soliton mass
$m$ (whose relation to $\mu_{bulk}$ is given by (\ref{eq:massgap})) are given by
\be
L = N \delta \,, \qquad m={2\over \delta} e^{-\pi \Lambda} \,,
\label{continuumlimit}
\ee
respectively. Changing to the rescaled rapidity variable
$\theta = \pi \lambda$, and setting $f^{(+)}(\theta) = 2\pi i
h^{(+)}(\theta)$, one arrives at the continuum NLIE for $f^{(+)}(\theta)$
\be
f^{(+)}(\theta) &=& 2i m L \sinh \theta + i P_{bdry}^{(+)}(\theta) 
+ i g(\theta) \non \\
&+& {2i\over \pi} \int_{-\infty}^{\infty} d\theta'\ \Im m\
G(\theta-\theta' - i \epsilon)\
\ln (1 - e^{f^{(+)}(\theta' + i \epsilon)}) \,,
\label{DDVeqn}
\ee
where $G(\theta)$ is given by
\be
G(\theta) =  {1\over 2\pi} \int_{-\infty}^{\infty} d\omega\ 
e^{-i \omega \theta/\pi} \hat G(\omega) \,,
\label{Gtheta}
\ee
and the Fourier transform $\hat G(\omega)$ is given by
\be
\hat G(\omega) =  {\sinh((\nu-2)\omega/2) \over 
2\sinh((\nu-1)\omega/2) \cosh(\omega/2)} \,.
\ee 
Furthermore, $P_{bdry}^{(+)}(\theta)$ is the odd function satisfying
${P_{bdry}^{(+)}}'(\theta) = 2 R^{(+)}(\theta)$, where $R^{(+)}(\theta)$ 
is given (as in (\ref{Gtheta})) in terms of its Fourier transform
\be
\lefteqn{\hat R^{(+)}(\omega) =   
{\sinh((\nu-2)\omega/4) \cosh(\nu \omega/4)\over 
\sinh((\nu-1)\omega/2) \cosh(\omega/2)} 
+ {\sinh((\nu-2)\omega/2) \over 
2\sinh((\nu-1)\omega/2) \cosh(\omega/2)}} \label{pbdry}\\
& & + {s_{+}\sinh((\nu-|2a_{+}-1|)\omega/2) \over 
2\sinh((\nu-1)\omega/2) \cosh(\omega/2)} + 
{\sinh((1+2ib_{+})\omega/2)\over 
2\sinh((\nu-1)\omega/2) \cosh(\omega/2)} 
+ (+ \leftrightarrow -) \non \,, 
\ee 
where $(+ \leftrightarrow -)$ is a shorthand for two additional terms
which are the same as those on the second line of (\ref{pbdry}), but
with $a_{+}, s_{+}$ and $b_{+}$ replaced by $a_{-}, s_{-}$ and $b_{-}$,
respectively. 

Moreover, $g(\theta)$ is the source term.  For simplicity, we
henceforth restrict our attention to source contributions from holes
and special roots; other bulk sources (close or wide roots) can
presumably be treated in the same manner as in \cite{DDV2, ABR}.  The
source term is therefore given by
\be
g(\theta) &=& \sum_{j=1}^{N_{H}} \left[ 
\chi(\theta - \theta_{j}^{H}) 
+ \chi(\theta + \theta_{j}^{H}) \right] \label{source} \\ 
&-& \sum_{j=1}^{N_{S}} \left[ 
  \chi(\theta - \theta_{j}^{S} + i\epsilon) 
+ \chi(\theta - \theta_{j}^{S} - i\epsilon)
+ \chi(\theta + \theta_{j}^{S} + i\epsilon) 
+ \chi(\theta + \theta_{j}^{S} - i\epsilon) \right] \non
\,,
\ee 
where $\chi(\theta)$ is the odd function satisfying $\chi'(\theta) = 2
G(\theta)$, the latter function being given by (\ref{Gtheta}).
Finally, $\theta_{j}^{H}$ and $\theta_{j}^{S}$ are the positions of
the holes and special roots, respectively, whose corresponding
distinct, positive integers we label by $I_{j}^{H}$ and $I_{j}^{S}$,
\be
f^{(+)}(\theta_{j}^{H}) = 2\pi i I_{j}^{H} \,, \qquad 
f^{(+)}(\theta_{j}^{S}) = 2\pi i I_{j}^{S} \,.
\label{sourceintegers}
\ee 
For the continuum model, the value $I_{j}=0$ is excluded because the 
corresponding rapidity $\theta_{j}=0$ is not physical.

The energy is given by
\be
E = \epsilon_{bulk} L + \epsilon_{boundary} + E_{Casimir} \,,
\ee
where $\epsilon_{bulk}$ and $\epsilon_{boundary}$ are given by \cite{AN}
\be
\epsilon_{bulk} = {1\over 4} m^{2}  \cot( \nu \pi/2)
\label{SGcontbulkenergy}
\ee
and
\be
\epsilon_{boundary} = - {m\over 2} \left[ -\cot(\nu \pi/4) -1 + 
{\cos((\nu -  2 s_{+} a_{+})\pi/2)\over \sin(\nu \pi/2)} +
{\cosh(\pi b_{+})\over \sin(\nu \pi/2)}  + (+ \leftrightarrow -) 
\right]
\,,
\label{SGcontboundenergy}
\ee
and $E_{Casimir}$ (order $1/L$) is given by
\be
E_{Casimir} &=& m \sum_{j=1}^{N_{H}} \cosh \theta_{j}^{H}
- m \sum_{j=1}^{N_{S}} \left[ 
\cosh (\theta_{j}^{S} + i\epsilon) +
\cosh (\theta_{j}^{S} - i\epsilon) \right] \non \\
&-& {m\over 2\pi} \int_{-\infty}^{\infty} d\theta\ 
\Im m\ 
\sinh (\theta+ i \epsilon)  
\ln (1 - e^{f^{(+)}(\theta + i \epsilon)})
\,.
\label{SGCasimir}
\ee

The following relations between the continuum IR ($\eta_{\pm} \,,
\vartheta_{\pm}$) and lattice ($a_{\pm} \,, b_{\pm}$) boundary
parameters were obtained in \cite{AN} \footnote{We have already
implicitly noted the relation between the continuum and lattice {\it bulk}
parameters in (\ref{bulkparamreltn}). Indeed, in \cite{AN}, $\nu$ is defined 
as $\nu \equiv \pi/\mu$, where $\mu \in (0 \,, \pi)$ is the bulk
anisotropy parameter of the XXZ spin chain. Here we do not explicitly 
introduce the lattice anisotropy parameter $\mu$ in an effort to
reduce the number of parameters appearing in the paper.}
\be
\eta_{\pm} &=& \mp {\pi( s_{\pm} \nu - 2 a_{\pm})\over 2(\nu-1)} \,, \non \\
\vartheta_{\pm} &=& {\pi b_{\pm}\over \nu-1}  \,.
\label{boundparamreltn1}
\ee
These relations were obtained by comparing the continuum \cite{BPT1,
AZ1} and lattice expressions for the boundary energy.
A relation between $\gamma_{\pm}$ and $c_{\pm}$ was also conjectured
in  \cite{AN}
\be
\gamma_{\pm} = {\pi c_{\pm}\over \nu-1} \,,
\label{cgamma}
\ee
for which we shall find further support from analysis of the UV and IR
limits in Sections \ref{subsec:UVNLIE} and \ref{subsec:IR} below.
Note that these relations together with the constraints
(\ref{latticeconstraints}) among the lattice boundary parameters imply
corresponding constraints among the continuum IR boundary parameters,
\be
\eta_{-} - \eta_{+} &=& \mp |\gamma_{-} - \gamma_{+}| 
+  {\pi\over \nu-1} \left[{1\over 2}(s_{+} + s_{-}) \nu - k \right]
\,, \non \\
\vartheta_{-} + \vartheta_{+} &=& 0 \,.
\label{continuumconstraints}
\ee
The UV and IR limits also imply that, in the continuum limit, the 
integer $k$ must be restricted to odd values,
\be
k = \mbox{ odd } \,.
\label{odd}
\ee 
The UV-IR relations (\ref{gammakappa}), (\ref{alyosha1}), (\ref{alyosha2}) imply 
corresponding constraints among the UV parameters appearing in the boundary 
sine-Gordon action.
We emphasize that our NLIE describes the sine-Gordon model only for 
values of boundary parameters which satisfy these constraints.

It is interesting to consider the continuum version of the lattice 
counting equation (\ref{latticecounteq}).
For the case of periodic boundary conditions, Destri and de Vega have 
argued \cite{DDV2} that the continuum result coincides with the 
lattice result without the integer part term. It is therefore plausible
that for the case at hand the continuum counting equation is given by
\be
N_{H} - 2N_{S} = M_{C} + 2M_{W}\mbox{step}(\nu-2)  
+1 +{1\over 2}(s_{+}+s_{-}) - k  \,.
\label{continuumcounteq}
\ee 
We shall find some support for this conjecture when we consider the
UV limit below. \footnote{We recall \cite{DDV2} that special 
objects cannot appear in the IR limit.  They can appear as $m L 
\rightarrow 0$, in such a way that $N_{H, eff}=N_{H} - 2N_{S}$ remains
constant.}

\subsection{Ultraviolet limit}\label{subsec:UVNLIE}

We now consider the UV limit $m L \rightarrow 0$. In this
limit, only large values of $|\theta|$ contribute to the Casimir 
energy.  \footnote{Since the counting function is odd, we consider
explicitly only $\theta \rightarrow \infty$, 
and we double the result in order to also account for the
$\theta \rightarrow -\infty$ contribution.}
Indeed, the driving term of the NLIE implies that one must consider
the scaling limit
\be
\theta =  \ln {1\over m L} + \tilde \theta \,,
\ee
where $\tilde \theta$ is finite as $m L \rightarrow 0$. Hence, one must
distinguish whether the positions of the sources remain finite
or scale in the same manner. Let $N_{H}^{0}$ ($N_{H}^{\infty}$) 
be the number of holes whose positions $\theta_{j}^{H\ 0}$ 
($\theta_{j}^{H\ \infty}$) remain finite (scale), 
with corresponding integers $I_{j}^{H\ 0}$ 
($I_{j}^{H\ \infty}$, respectively); 
and similarly for the special roots.  Hence,
\be
\theta_{j}^{H\ \infty} =  \ln {1\over m L} 
+ \tilde \theta_{j}^{H\ \infty} \,, \qquad 
\theta_{j}^{S\ \infty} =  \ln {1\over m L} 
+ \tilde \theta_{j}^{S\ \infty} \,,
\ee 
with $\tilde \theta_{j}^{H\ \infty}$ and  $\tilde \theta_{j}^{S\ 
\infty}$ finite as $m L \rightarrow 0$, and
\be
N_{H} = N_{H}^{0} + N_{H}^{\infty} \,, \qquad 
N_{S} = N_{S}^{0} + N_{S}^{\infty} \,.
\label{evident}
\ee
Proceeding to compute the Casimir energy
from (\ref{DDVeqn}) and (\ref{SGCasimir}) as in 
\cite{DDV2}-\cite{LMSS}, imposing the constraints
(\ref{latticeconstraints}), and
recalling that $E_{Casimir} = -  {c_{eff} \pi\over 24 L}$ with 
$c_{eff} = 1 -24\Delta_{n}$, we obtain 
\be
\Delta_{n} &=& 
{1\over 4 \nu (\nu-1)} \Bigg\{  \nu \left[
{1\over 2}(s_{+} + s_{-}) + N_{H}^{0} - N_{H}^{\infty}
-2N_{S}^{0} + 2N_{S}^{\infty} \right]
-\left(k-1+  2N_{H}^{0}-4N_{S}^{0}\right) \non \\
&\mp& |c_{-}-c_{+}| \Bigg\}^{2} 
 + \sum_{j=1}^{N_{H}^{\infty}} I_{j}^{H\ \infty} 
 -2\sum_{j=1}^{N_{S}^{\infty}} I_{j}^{S\ \infty}
 -{1\over 2}(N_{H}^{\infty} -2 N_{S}^{\infty})
 (N_{H}^{\infty} -2 N_{S}^{\infty}+1) \,.
\label{DeltaNLIE}
\ee

The result (\ref{DeltaNLIE}) for the conformal dimension is consistent
with the CFT result (\ref{DeltaCFT}) if we identify 
\be
c_{\pm} = {4\over \beta}(\nu-1)\kappa_{\pm} \,,
\label{ckappa}
\ee
and set
\be
{1\over 2}(s_{+} + s_{-}) + N_{H}^{0} - N_{H}^{\infty}
-2N_{S}^{0} + 2N_{S}^{\infty} = k-1+  2N_{H}^{0}-4N_{S}^{0} = 2n \,,
\label{comparison}
\ee
where $n$ is an integer.  The relation (\ref{ckappa}) together with
(\ref{gammakappa}) implies the relation (\ref{cgamma}) between the
boundary parameters $\gamma_{\pm}$ and $c_{\pm}$.  \footnote{We remark
that (\ref{ckappa}) and the periodicity (\ref{kappaperiodicity}) imply
the periodicity $c_{\pm} \rightarrow c_{\pm} + 2(\nu-1)$; i.e., the
periodicity $c_{\pm} \rightarrow c_{\pm} + 2\nu$ of the lattice model
\cite{AN} becomes ``renormalized'' in the continuum.} Note that the
result  (\ref{DeltaNLIE}) for the conformal dimension contains not
only the contribution  (\ref{DeltaCFT}) corresponding to the highest 
weights, but also additional integer-valued terms corresponding to
their descendants.

It follows from (\ref{evident}) and (\ref{comparison}) that 
\be
N_{H} - 2N_{S} = 1 +{1\over 2}(s_{+}+s_{-}) - k  \,,
\label{continuumcounteqspecial}
\ee
which is in agreement with the conjectured continuum counting 
equation (\ref{continuumcounteq}) for the special case $M_{C}=M_{W}=0$
that we are considering. Moreover, (\ref{comparison}) implies the 
result (\ref{odd}) that $k$ must be an odd integer. 

Here are three examples:
\begin{description}
    \item[Ex. 1] A Bethe state consisting of only real roots and no holes or
    special roots ($N_{H} = N_{S} = 0$) with $k=1$ has, according to
    (\ref{continuumcounteqspecial}), a ``good'' UV limit only if the
    boundary parameters satisfy $s_{+}+s_{-}=0$; and its UV dimension,
    according to (\ref{comparison}), is given by (\ref{DeltaCFT}) with
    $n=0$.
    \item[Ex. 2] A Bethe state consisting of only real roots and 1 hole and
    no special roots ($N_{H} = 1\,, N_{S} = 0$) with $k=1$ has a
    ``good'' UV limit only if $s_{+}+s_{-}=2$.  Its UV dimension is
    given by (\ref{DeltaCFT}) with either $n=1$ or $n=0$, depending on
    whether in the UV limit the hole's rapidity remains finite
    ($N_{H}^{0}=1\,, N_{H}^{\infty}=0$) or infinite ($N_{H}^{0}=0\,,
    N_{H}^{\infty}=1$), respectively.
    \item[Ex. 3] For $k=-1$, it is possible to have a Bethe state with a
    ``good'' UV limit consisting of only real roots and 2 holes and no
    special roots ($N_{H} = 2\,, N_{S} = 0$), provided
    $s_{+}+s_{-}=0$.  Its UV dimension is given by (\ref{DeltaCFT})
    with $n$ values equal to either $1\,, 0\,,$ or $-1$ depending on
    whether in the UV limit the hole rapidities are either both
    finite, one finite and one infinite, or both infinite,
    respectively.
 \end{description}

\subsection{Infrared limit}\label{subsec:IR}

We verify in this Section that the IR limit of the NLIE
for a one-hole state is equivalent to the Yang equation for a particle
on an interval. Indeed, in the IR limit $m L \rightarrow \infty$, 
the integral terms in the
NLIE (\ref{DDVeqn}) and in the energy formula (\ref{SGCasimir}) are of
order $O(e^{-m L})$ and can therefore be neglected.  For a single hole
with rapidity $\theta_{H}$, the NLIE becomes
\be
f^{(+)}(\theta) = 2i m L \sinh \theta + i P_{bdry}^{(+)}(\theta) 
+ i\chi(\theta-\theta_{H}) + i\chi(\theta+\theta_{H}) 
\,.
\ee
Noting that $e^{f^{(+)}(\theta_{H})}=1$ on account of Eq.
(\ref{sourceintegers}), we obtain the following relation for
$\theta_{H}$
\be
e^{2i m L \sinh \theta_{H}} 
e^{i\left( P_{bdry}^{(+)}(\theta_{H}) + \chi(2\theta_{H}) \right)} = 1 \,.
\label{NLIEquantization}
\ee
A similar relation can be derived from the negated counting function,
\be
e^{2i m L \sinh \theta_{H}} 
e^{i\left( P_{bdry}^{(-)}(\theta_{H}) + \chi(2\theta_{H}) \right)} = 1 \,,
\ee
where $P_{bdry}^{(-)}(\theta)$ differs from $P_{bdry}^{(+)}(\theta)$ 
by the negation of the boundary parameters (\ref{negated}).
These relations should be equivalent to the Yang
equation for a particle on an interval of length $L$,
\be
e^{2i m L \sinh \theta_{H}} 
R(\theta_{H}\,; \eta_{+}, \vartheta_{+}, \gamma_{+}) 
R(\theta_{H}\,; \eta_{-}, \vartheta_{-}, \gamma_{-}) 
|\theta_{H}, (\pm) \rangle 
= |\theta_{H}, (\pm)  \rangle \,,
\label{yangequation}
\ee
where the boundary $S$ matrices $R(\theta\,; \eta_{\pm},
\vartheta_{\pm}, \gamma_{\pm})$ are given by (\ref{boundSmatrix}),
and $|\theta_{H}, (\pm) \rangle$ denote the 
two possible one-particle states. \footnote{In the case of Dirichlet 
boundary conditions, $|\theta_{H}, (+) \rangle$ and $|\theta_{H}, (-) 
\rangle$ would correspond to one-soliton and one-antisoliton states, 
respectively. For this case, a similar approach for computing boundary
$S$ matrices was considered in \cite{DMN}.}

In other words, dropping the subscript $H$ of the hole rapidity, the
expressions $e^{i\left( P_{bdry}^{(\pm)}(\theta) + \chi(2\theta) \right)}$
should be equal to the two eigenvalues of the Yang matrix $Y(\theta)$, 
which is defined by
\be
Y(\theta) = R(\theta\,; \eta_{+}, \vartheta_{+}, \gamma_{+}) 
R(\theta\,; \eta_{-}, \vartheta_{-}, \gamma_{-}) \,.
\label{yangmatrix}
\ee 

Indeed, for a state with real roots and one hole, Eq.
(\ref{continuumcounteq}) implies that $k={1\over 2}(s_{+}+s_{-})$.
Since $k$ must be odd, it follows that the only two possibilities are
$k=s_{\pm}=1$ or $k=s_{\pm}=-1$.  For definiteness, we consider the
former case,
$k=s_{\pm}=1$, and therefore, ${1\over 2} < a_{\pm} < {1\over 2} + \nu$.  
From the definitions
of $P_{bdry}^{(\pm)}(\theta)$ and $\chi(\theta)$, it follows that
\be
P_{bdry}^{(\pm)}(\theta) + \chi(2\theta) = 2 \int_{0}^{\infty}
{d\omega\over \omega} 
\left[\sin (\omega \theta/\pi) \hat R^{(\pm)}(\omega) +
\sin (2\omega \theta/\pi) \hat G(\omega) \right] \,.
\ee 
With the help of the expression (\ref{pbdry}) for $\hat
R^{(+)}(\omega)$ (and a similar expression with the
boundary parameters negated (\ref{negated}) for $\hat
R^{(-)}(\omega)$) and the identity \cite{FS}
\be
\lefteqn{{\sinh((\nu-2)\omega/4) \cosh(\nu \omega/4)\over 
\sinh((\nu-1)\omega/2) \cosh(\omega/2)} 
+ {\sinh((\nu-2)\omega/2) \over 
2\sinh((\nu-1)\omega/2) \cosh(\omega/2)}} \non \\
& & = {2\sinh((\nu-2)\omega/4) \sinh(3\omega/4)\over 
\sinh((\nu-1)\omega/4) \sinh \omega} - \hat G(\omega/2) \,,
\ee 
we obtain 
\be
e^{i\left( P_{bdry}^{(\pm)}(\theta) + \chi(2\theta) \right)}
&=& \exp 2i \int_{0}^{\infty}
{d\omega\over \omega} \sin (2\omega \theta/\pi) \Big\{
{2\sinh((\nu-2)\omega/2) \sinh(3\omega/2)\over 
\sinh((\nu-1)\omega/2) \sinh (2\omega)}  \\
&+& {\sinh((1 \pm (\nu - 2a_{+}))\omega) \over 
2\sinh((\nu-1)\omega) \cosh \omega} + 
{\sinh((1 \pm 2ib_{+})\omega) \over 
2\sinh((\nu-1)\omega) \cosh \omega}
+ (+ \leftrightarrow -) \Big\} \,. \non
\ee
Finally, using the relations (\ref{boundparamreltn1}) between the
lattice and continuum IR boundary parameters, we obtain
\be
e^{i\left( P_{bdry}^{(\pm)}(\theta) + \chi(2\theta) \right)}
&=& \exp 2i \int_{0}^{\infty}
{d\omega\over \omega} \sin (2\omega \theta/\pi) \Big\{
{2\sinh((\nu-2)\omega/2) \sinh(3\omega/2)\over 
\sinh((\nu-1)\omega/2) \sinh (2\omega)}  \\
&+& {\sinh((1 \mp (\nu-1) 2\eta_{+}/\pi)\omega) \over 
2\sinh((\nu-1)\omega) \cosh \omega} 
+ {\sinh((1 \pm (\nu-1) 2\eta_{-}/\pi)\omega) \over 
2\sinh((\nu-1)\omega) \cosh \omega} \non \\
&+&{\sinh((1 \pm (\nu-1) 2i\vartheta_{+}/\pi)\omega) \over 
2\sinh((\nu-1)\omega) \cosh \omega}
+{\sinh((1 \pm (\nu-1) 2i\vartheta_{-}/\pi)\omega) \over 
2\sinh((\nu-1)\omega) \cosh \omega} \Big\} \,. \non
\label{IRNLIE}
\ee

We now turn to the computation of the eigenvalues of the Yang
matrix (\ref{yangmatrix}). Recalling (\ref{boundSmatrix}), we see
that the eigenvalues $y^{(\pm)}$ of $Y(\theta)$ are given by
\be
y^{(\pm)} = r_{0}(\theta)^{2} r_{1}(\theta\,; \eta_{+}, \vartheta_{+})
r_{1}(\theta\,; \eta_{-}, \vartheta_{-}) \Lambda^{(\pm)} \,,
\ee
where $\Lambda^{(\pm)}$ denote the 
eigenvalues of $M(\theta\,; \eta_{+}, \vartheta_{+}, \gamma_{+}) 
M(\theta\,; \eta_{-}, \vartheta_{-}, \gamma_{-})$.
Although the expressions for $\Lambda^{(\pm)}$ are generally  
very complicated, a remarkable simplification occurs if the boundary 
parameters satisfy the constraints (\ref{continuumconstraints}),
(\ref{odd}). Indeed, in that case, the eigenvalues
are factorizable into a product of trigonometric functions,
\be
\Lambda^{(\pm)} = \cos(-\eta_{+} \mp i \lambda \theta) 
\cos(\eta_{-} \mp i \lambda \theta) 
\cos(i\vartheta_{+} \mp i\lambda \theta)
\cos(i\vartheta_{-} \mp i\lambda \theta) \,.
\ee 
Recalling the expressions (\ref{r0r1}) for $r_{0}(\theta)$ and 
$r_{1}(\theta\,; \eta, \vartheta)$, using the identity
\be
{1\over \cos x} \cos(x \mp i \lambda \theta)\ \sigma(x, \theta)
= \exp \left\{ 2i\int_{0}^{\infty} {d\omega\over \omega}
\sin (2\theta \omega/ \pi) 
{\sinh( (1 \pm (\nu-1) 2 x/ \pi) \omega)\over
2\sinh((\nu-1) \omega) \cosh \omega} \right\} \,,
\ee
and comparing with (\ref{IRNLIE}), we obtain the desired result
\be
e^{i\left( P_{bdry}^{(\pm)}(\theta) + \chi(2\theta) \right)}
= r_{0}(\theta)^{2} r_{1}(\theta\,; \eta_{+}, \vartheta_{+})
r_{1}(\theta\,; \eta_{-}, \vartheta_{-}) \Lambda^{(\pm)} = y^{(\pm)} \,.
\ee
That is, we have verified that the IR limit of the NLIE for a one-hole
state is equivalent to the Yang equation for a particle on an
interval.  We stress that the boundary $S$ matrices entering the Yang
equation are not diagonal.

We remark that, for the case of $N_{H}$ holes, the Casimir energy in
the IR limit becomes
\be
E_{Casimir} \rightarrow m N_{H} \,.
\label{CasimirIR}
\ee
Indeed, as already noted, the integral term in the energy formula
(\ref{SGCasimir}) can be neglected; thus, only the first term of that
formula survives.  Moreover, the hole rapidities go as $\theta_{j}^{H}
\sim {1\over m L}$ (since $m L \sinh \theta_{j}^{H} \sim 1$) for $L
\rightarrow \infty$, which leads to the result (\ref{CasimirIR}).

\section{Numerical results}\label{sec:numerical}

The NLIE (\ref{DDVeqn}) can be solved numerically by iteration, and
the corresponding Casimir energy can then be evaluated with
(\ref{SGCasimir}).  It is not evident how to best present such
numerical results for the full range of $L \in (0 \,, \infty)$.  The
difficulty is that, in the UV limit ($L \rightarrow 0$), $E_{Casimir}
= - {\pi c_{eff}\over 24 L}$ diverges and $c_{eff}$ is finite; while
in the IR limit ($L \rightarrow \infty$), the reverse is true:
$E_{Casimir}$ is finite (\ref{CasimirIR}) and $c_{eff}$ diverges (if
the number of holes is not zero).  That is, neither $E_{Casimir}$ nor
$c_{eff}$ remain finite over the full range of $L$.  Following
\cite{PCA}, we consider the dimensionless quantity (``normalized
energy'')
\be
{\cal E}={L E_{Casimir}\over \pi + m L}  = 
- {\pi c_{eff}\over 24 \left(\pi + m L\right)}
\,,
\ee 
whose UV and IR limits are both finite: 
\be
{\cal E} &\rightarrow &
\Delta_{n} - {1\over 24} \quad 
\mbox{ for  } \quad  L \rightarrow 0 \,, \label{calElimitsUV} \\
{\cal E}  &\rightarrow & N_{H} \quad  \quad  \quad 
\mbox{ for  } \quad L \rightarrow \infty
\,. \label{calElimitsIR}
\ee 

We have plotted ${\cal E}$ as a function of $\ln l$, where $l \equiv m
L$ is the dimensionless scale parameter, for various states.
\footnote{For the case of 0 holes, we plot $-c_{eff}/24$ vs.  $\ln l$,
which also has the limiting values (\ref{calElimitsUV}),
(\ref{calElimitsIR})  with $n=N_{H}=0$.}

\subsection{Ground state}\label{subsec:ground}

\begin{figure}[tb]
	\centering
	\includegraphics[width=0.80\textwidth]{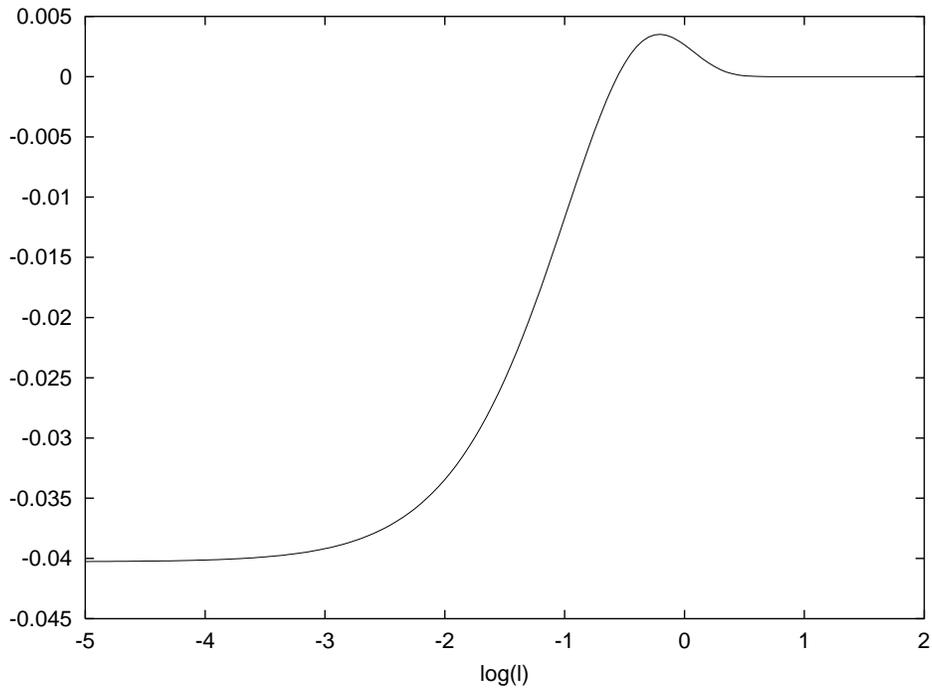}
	\caption[xxx]{\parbox[t]{0.92\textwidth}{
	$-c_{eff}/24$ vs. $\log l$ for ground state (0 holes),
	with parameter values $\nu=1.93$, $a_{+}=1.8$, $a_{-}= -0.9$, 
	and $b_{+}=-b_{-}=0.41444$. See Ex. 1 in Sec. \ref{subsec:UVNLIE},
	and Sec. \ref{subsec:ground}.}
	}
	\label{fig:ground}
\end{figure}

NLIE results for the ground state (0 holes), whose UV limit is
discussed in Ex.  1 at the end of Section \ref{subsec:UVNLIE}, are
presented in Fig.  \ref{fig:ground}.  As expected, the value of $-c_{eff}/24$ in the IR
limit is 0; and in the UV limit agrees well with the analytical
result for $\Delta_{0}$ given by (\ref{DeltaCFT}), (\ref{DeltaNLIE}).

We define three regions of $l = m L$ in which we further test, with
different methods, the ground state energy level obtained by
numerically solving the NLIE:

\begin{itemize}
\item The UV region is the small volume region, $l<10^{-1}$; here we compare
it with boundary conformal perturbation theory (BCPT). 
\item In the intermediate region, $l\sim 1$, we test it against 
truncated conformal space approach (TCSA) \cite{YZ,DPTW}. 
\item In the IR region, where the volume is large, $l>10$, we compute its
L\"uscher-type \cite{Lu,BPT3} correction. 
\end{itemize}
In all regions we obtain a perfect confirmation of the correctness
of our NLIE.

\subsubsection{UV region}\label{subsubsec:uv}

Combining the formulae from Appendix A which describe the BCPT and NLIE
schemes, we obtain the small volume expansion of the NLIE ground state energy
\be 
E_{NLIE}(L) &=& -\epsilon_{bulk}L-\epsilon_{boundary}
+\frac{\pi}{L}\Bigg(E_{|0\rangle}-\frac{1}{24} \non \\
&+& c_{2}^{0}\left(\frac{\pi}{L}\right)^{2(\Delta-1)}
+c_{4}^{0}\left(\frac{\pi}{L}\right)^{4(\Delta-1)}
+c_{6}^{0}\left(\frac{\pi}{L}\right)^{6(\Delta-1)}
+\dots\Bigg) \,. \label{smallvol}
\ee 
Note that $E_{|0\rangle}$ is the
conformal dimension of the ground state, which is given by
(\ref{DeltaCFT}) with $n=0$; that is, $E_{|0\rangle} = {\kappa^{2}\over
2\pi}$. Also, as in Sec. \ref{subsec:bsg}, here
$\Delta = {\beta^{2}\over 8\pi}= {\nu-1\over \nu}$. The bulk and
boundary energies are given by Eqs. (\ref{SGcontbulkenergy}) and  
(\ref{SGcontboundenergy}), respectively.
Computing numerically the ground-state energy for small volumes, the coefficients
$c_{2}^{0}\ m^{2(\Delta -1)}$ can be extracted. 
Table \ref{c2table} shows a comparison
between the numerically measured coefficients \footnote{Specifically, 
we computed $m^{-1} E_{NLIE}$ for 100 values of $l$, 
from $l=10^{-5}$ to $l=10^{-3}$, which we fitted to the curve
(\ref{smallvol}) to obtain estimates for $c_{j}^{0}\ m^{j(\Delta -1)}$.} 
and the exact values calculated from BCPT  (\ref{eq:finalresult}),
(\ref{eq:massgap}) for  various values of the bulk coupling constant $\nu$ and
for the same values of boundary parameters used
to generate Fig. \ref{fig:ground}. The agreement is convincing and is of the order of our
numerical precision.

\begin{table}[htb] 
  \centering
  \begin{tabular}{|c|c|c|}\hline
    $\nu$ & NLIE $c_{2}^{0}\ m^{2(\Delta -1)}$  & BCPT $c_{2}^{0}\ m^{2(\Delta -1)}$\\
    \hline
    1.70     & -5.3215286975      & -5.3215288274 \\ 
    1.80     & -7.4632436186      & -7.4632435914 \\
    1.93     &-19.5148929102      &-19.5148929079 \\
    2.20     &  5.6819407377      &  5.6819407318 \\
    2.40     &  2.4879276564      &  2.4879276494 \\
    2.60     &  1.4601870563      &  1.4601870411 \\ 
    \hline
   \end{tabular}
   \caption{Comparison of NLIE and BCPT results for $c_{2}^{0}\ m^{2(\Delta -1)}$, 
   for various values of bulk coupling constant $\nu$ and for boundary parameter values 
   $a_{+}=1.8$, $a_{-}= -0.9$, and $b_{+}=-b_{-}=0.41444$.}
  \label{c2table}
\end{table}

\subsubsection{Intermediate region}\label{subsubsec:intermed}

In this region, the energy levels are not dominated only by the first
few terms in the UV expansion; instead, all the higher-order terms contribute
the same way. That is, a non-perturbative check is necessary. This
is provided by a TCSA calculation, which -- being a variational 
method -- sums up the perturbative series, in which all the coefficients are
calculated approximately in a finite-dimensional, truncated Hilbert
space. The difficulty is in the comparison. TCSA works if the dimension
of the perturbing operator is small, that is when $\nu$ is close
to one. In this domain, however, the NLIE is not convergent. So one
has to find a proper range, where the NLIE is convergent and the TCSA
is reliable enough. In Table \ref{TCSAtable} we present results 
for $\nu=1.2$ and for
boundary parameter values $a_{+}=1.2\,,a_{-}=-0.2$ and $b_{+}=b_{-}=0$.
The dimensionless NLIE ground state energy data are transformed into
the TCSA scheme by (\ref{eq:corresp}) and are compared to the
dimensionless TCSA data for different truncation levels and
dimensionless volumes, $l=mL$.

\begin{table}[htb] 
  \centering
    \begin{tabular}{|c|c|c|c|c|}
\hline 
Volume&
$l=0.7$&
$l=0.9$&
$l=1.1$&
$l=1.3$\tabularnewline
\hline
\hline 
$m^{-1}E_{TCSA}$~ with $E_{cut}=10$&
-0.32803&
-0.31248&
-0.31030&
-0.31498\tabularnewline
\hline 
$m^{-1}E_{TCSA}$~with $E_{cut}=12$&
-0.32834&
-0.31284&
-0.31072&
-0.31544\tabularnewline
\hline 
$m^{-1}E_{TCSA}$~with $E_{cut}=14$&
-0.32857&
-0.31311&
-0.31103&
-0.31579\tabularnewline
\hline 
$m^{-1}E_{TCSA}$~with $E_{cut}=16$&
-0.32875&
-0.31332&
-0.31127&
-0.31606\tabularnewline
\hline 
$m^{-1}(E_{NLIE}+\epsilon_{bulk}L+\epsilon_{bdry})$&
-0.33067&
-0.31559&
-0.31386&
-0.31895\tabularnewline
\hline
\end{tabular}
\caption{Comparison of NLIE and TCSA results for the ground-state energy, 
   for $\nu=1.2$ and for boundary parameter values 
   $a_{+}=1.2$, $a_{-}= -0.2$, and $b_{+}= b_{-}=0$.}
  \label{TCSAtable}
\end{table}
We can see that as we increase $E_{cut}$ the TCSA energies approach
the NLIE energy from above as a consequence of the variational nature
of the TCSA. The truncated Hilbert space with $E_{cut}=16$ contains
$6133$ states.

\subsubsection{IR region}\label{subsubsec:ir}

Here we check the exponentially small correction to the ground state
energy for large but finite volumes. This is dominated by the first
breather, with mass $m_{1}=2m\sin(\frac{\pi}{2\lambda})$, and is
given by \cite{BPT3} as 
\be 
E_{NLIE}\left(L\right)= m_{1}
\frac{1+\cos\frac{\pi}{2\lambda}-\sin\frac{\pi}{2\lambda}}
{1-\cos\frac{\pi}{2\lambda}+\sin\frac{\pi}{2\lambda}}
\tan\frac{\eta_{+}}{2\lambda}
\tanh\frac{\vartheta_{+}}{2\lambda}
\tan\frac{\eta_{-}}{2\lambda}
\tanh\frac{\vartheta_{-}}{2\lambda}
\mathrm{e}^{-m_{1}L}+\dots
\label{luschercor}
\ee 

\begin{figure}[htb]
	\centering
	\includegraphics[%
	  width=0.80\textwidth]{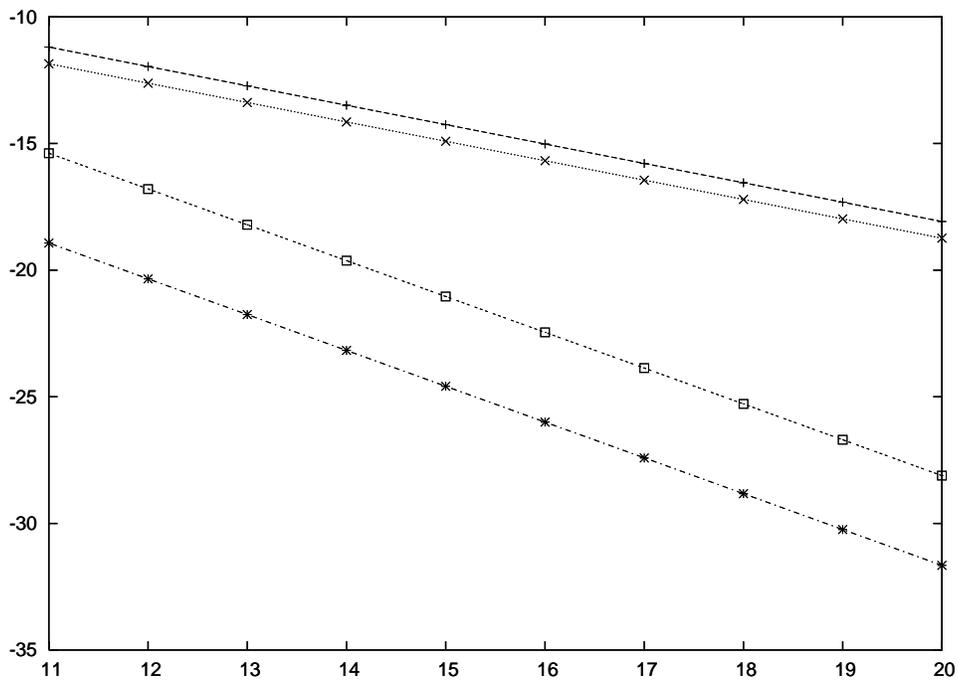}
	  \caption[xxx]{\parbox[t]{0.92\textwidth}{
	  $\log (m^{-1} E)$ vs. $l$ for the ground state, with
	  various values of the bulk and boundary parameters.	  
	  See Sec. \ref{subsubsec:ir}.}
		  }
		  \label{fig:Luscher}
\end{figure}

In Figure \ref{fig:Luscher} this correction is checked as a function of
$\nu = 1 + 1/\lambda$ and of the boundary parameters.
On the figure the logarithm of the dimensionless ground state energy
is plotted against the dimensionless volume. The upper two lines in
descending order are the L\"uscher corrections (\ref{luschercor}) 
for $\nu=1.25,\: b_{+}=-b_{-}=0.1$; and with
$a_{+}=1.2\:,\, a_{-}=-0.2$ for the first line, and $a_{+}=1.1\:,\, a_{-}=-0.3$
for the second line. The lower two lines have parameters 
$\nu=1.5,\: a_{+}=1.2\:,\, a_{-}=-0.2$; and with
$b_{+}=-b_{-}=1$ for the first line, while $b_{+}=-b_{-}=0.1$ for the
second line. The various boxes are the data of the numerical solution
of the NLIE. The agreement is really excellent.

\subsection{Excited states}\label{subsec:excited}

\begin{figure}[htb]
	\centering
	\includegraphics[width=0.80\textwidth]{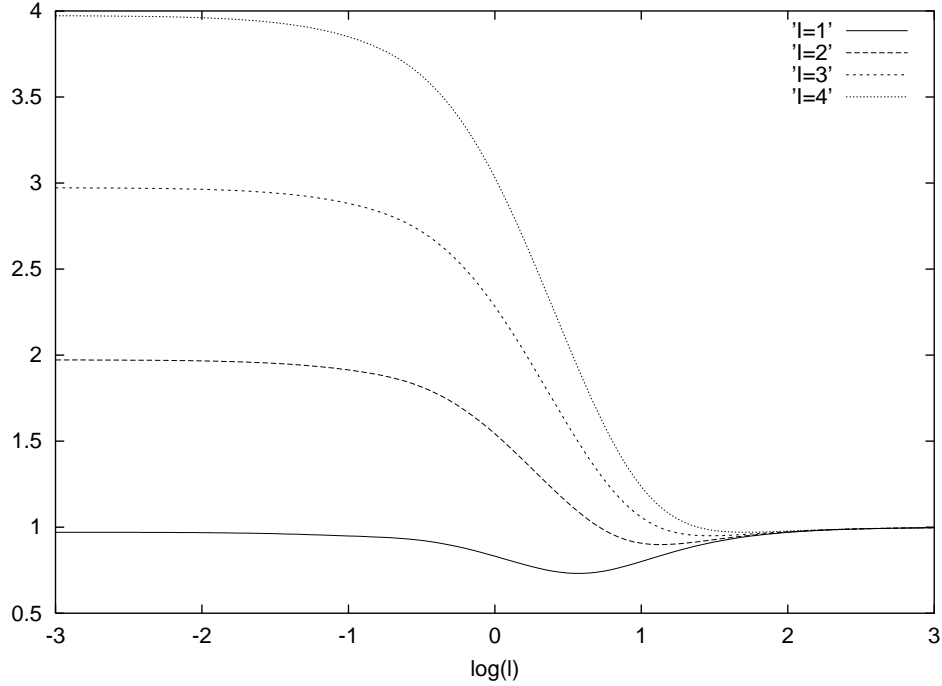}
	\caption[xxx]{\parbox[t]{1.0\textwidth}{
	${\cal E}$ vs. $\log l$ for 1-hole states
	with integer values $I^{H}=1, 2, 3, 4$, and
	with parameter values  $\nu=1.93$, $a_{+}=1.8$, $a_{-}= 1.9$, 
	and $b_{+}=-b_{-}=0.41444$. See Ex. 2 in Sec. \ref{subsec:UVNLIE},
	and Sec. \ref{subsec:excited}.}
	}
	\label{fig:onehole}
\end{figure}

\begin{figure}[htb]
	\centering
	\includegraphics[width=0.80\textwidth]{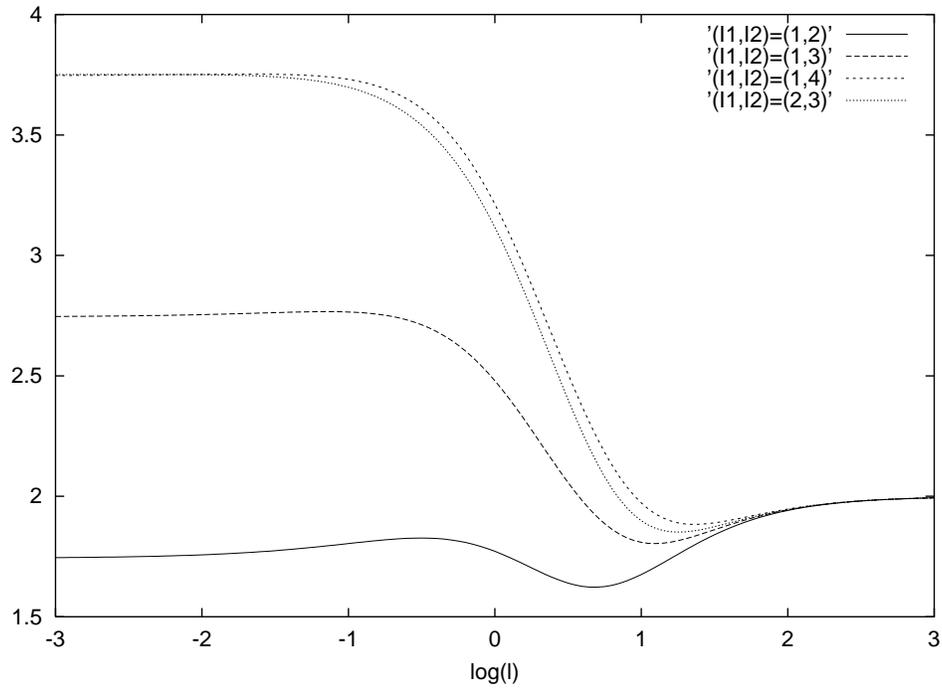}
	\caption[xxx]{\parbox[t]{1.0\textwidth}{
	${\cal E}$ vs. $\log l$ for 2-hole states
	with integer values $(I_{1}^{H}, I_{2}^{H})= (1, 2), 
	(1, 3), (1, 4), (2, 3)$, 
	and with parameter values  $\nu=2.13$, $a_{+}=1.8$, $a_{-}= -0.9$, 
	and $b_{+}=-b_{-}=0.50357$. See Ex. 3 in Sec. 
	\ref{subsec:UVNLIE}, and Sec. \ref{subsec:excited}.}
	}
	\label{fig:twohole}
\end{figure}

NLIE results for states with 1 and 2 holes, whose UV limits are
discussed in Exs.  2 and 3 at the end of Section \ref{subsec:UVNLIE},
are presented in Figs.  \ref{fig:onehole} and  \ref{fig:twohole}, respectively.  
Note that the values of
${\cal E}$ in the IR limit are 1 and 2, respectively, in agreement
with (\ref{calElimitsIR}).  Moreover, the values of ${\cal E}$ in the UV
limit agree well with (\ref{calElimitsUV}) and with the analytical
results for $\Delta_{n}$ given by (\ref{DeltaCFT}), (\ref{DeltaNLIE}).

In particular, for the 1-hole states (Fig.  \ref{fig:onehole}), we consider integer
values $I^{H} = 1, 2, 3, 4$; we find that the corresponding hole
rapidities $\theta^{H}$ become infinite in the UV limit, and thus, all
of these states have $n=0$.  The values of ${\cal E}$ in the UV limit
are spaced by 1 on account of the additional integer contribution to
$\Delta_{n}$ in (\ref{DeltaNLIE}): as $I^{H} = I^{H\ \infty}$ increases by
1, so does $\Delta_{n}$.  Similarly, for the 2-hole states (Fig.  \ref{fig:twohole}), 
we consider integer values $(I_{1}^{H}, I_{2}^{H})= (1, 2), (1, 3), (1,
4), (2, 3)$; we find that both hole rapidities become infinite in the
UV limit, and thus, the states have $n=-1$.  As $I_{1}^{H}+ I_{2}^{H}$
increases by 1, so does the limiting UV value of $\Delta_{n}$.  Hence, the
lowest line is $(I_{1}^{H}, I_{2}^{H})= (1, 2)$, the second-lowest
line is $(1,3)$, and the next two (almost degenerate) are $(2,3)$ and
$(1,4)$.

\section{Conclusion}\label{sec:conclude}

Starting from the Bethe Ansatz solution \cite{Ne, CLSW, NR} of the XXZ
model with general boundary terms, we have derived a nonlinear
integral equation for some bulk excited states of the
sine-Gordon model on a finite interval with general integrable
boundary interactions \cite{Sk1, GZ}, including boundary terms
proportional to $\partial_{y}\varphi$.  We have used this NLIE to
compute numerically the dimensions of these states as a function of
scale, and have checked the UV and IR limits analytically.
We have also verified that the ground-state NLIE agrees well with
boundary conformal perturbation theory (BCPT), boundary truncated
conformal space approach (BTCSA) and the boundary L\"uscher formula. 
An advantage of the latter approaches is that they are not restricted
to values of the boundary parameters that obey the constraints
(\ref{latticeconstraints}), (\ref{continuumconstraints}).
The consistency of the results provides support for the proposed
relations between the lattice and continuum boundary parameters.
%, as well as for the over-all approach.

The result (\ref{DeltaCFT}) for the conformal dimensions of a free
massless Boson with Neumann boundary conditions and
$\partial_{y}\varphi$ boundary terms, which is ``dual'' to the
corresponding result for a massless Boson with Dirichlet boundary
conditions \cite{Sa, AOS}, may have applications in other contexts,
such as string theory.

There are many issues that remain to be addressed.  Among these are
the proper treatment of complex (bulk) and imaginary (boundary)
sources in the NLIE. While the former problem is in principle
understood \cite{DDV2, ABR}, the latter problem is still not well
understood even in the simpler Dirichlet case \cite{ABR, SS}.
Moreover, it would be interesting to extend the comparison of NLIE
with BCPT and BTCSA also to excited states.

\section*{Acknowledgments}

We are grateful to F. Ravanini for valuable discussions.
This work was supported in part by the Korea Research Foundation
2002-070-C00025 (C.A.); 
by the EC network {}``EUCLID'', contract number HPRN-CT-2002-00325, 
and Hungarian research funds OTKA D42209, T037674, T034299 and
T043582 (Z.B., L.P. and G.T.);
and by
the National Science Foundation under Grants PHY-0098088 and
PHY-0244261, and by a UM Provost Award (R.N.).

\appendix

\section{Boundary Conformal Perturbation Theory \\
and Boundary Truncated Conformal Space Approach}

Boundary conformal perturbation theory (BCPT) and boundary truncated
conformal space approach (BTCSA) \cite{DPTW, BPT2} can be applied
if the theory is a relevant perturbation of a boundary conformal field
theory: 
\[
L=L_{BCFT}+L_{pert}=L_{BCFT}-\mu_{bulk}\int_{x_{-}}^{x^{+}}\Phi(x,t)dx
-\mu_{-}\Psi_{-}(t)-\mu_{+}\Psi_{+}(t)\quad,\]
where $L_{BCFT}$ is the Lagrangian of the UV limiting boundary conformal
field theory, $\Phi(x,t)$ is a relevant bulk primary field of weights
$(h,\bar{h})$ and $\Psi_{\pm}$ are relevant boundary fields living
on the left/right boundaries of the strip with weights $\Delta_{\pm}$.
For simplicity we will suppose that $h=\bar{h}=\Delta_{-}=\Delta_{+}=:\Delta.$
and put $x_{-}=0;\; x_{+}=L$. \footnote{We emphasize that 
Secs. \ref{subsec:hamiltonian} and \ref{subsec:lagrangian} are more general 
than the main body of the text, as they are valid for any perturbed boundary 
conformal field theory.}

\subsection{Hamiltonian approach}\label{subsec:hamiltonian}

We are interested in the spectrum of the Hamiltonian:
\[
H(L)=H_{BCFT}(L)+\mu_{bulk}\int_{0}^{L}\Phi(x,t)dx
+\mu_{-}\Psi_{-}(t)+\mu_{+}\Psi_{+}(t)\quad.\]
The volume dependence can be obtained by mapping the system to the
upper half plane (UHP) via $z=e^{i\frac{\pi}{L}(x+t)}$, where $t=-iy$
is the Euclidean time. The transformation rules of the primary fields
are given by: 
\begin{equation}
\Phi(x,t)=\left(\frac{\pi^{2}}{L^{2}}z\bar{z}\right)^{\Delta}\Phi(z,\bar{z})\quad;
\quad\Psi_{\pm}(t)=\left(\mp\frac{\pi}{L}z\right)^{\Delta}\Psi(z)\quad,
\label{eq:trafo}
\end{equation}
 Changing the integration variable to $\theta=\frac{\pi}{L}x$ and
taking the Hamiltonian at $t=0$, we have:
\begin{eqnarray*}
H(L) &=& \frac{\pi}{L}\left(L_{0}-\frac{c}{24}\right)
+\mu_{bulk}\left(\frac{\pi}{L}\right)^{2\Delta-1}\int_{0}^{\pi}
\Phi(e^{i\theta},e^{-i\theta})d\theta \\
&+& \mu_{-}\left(\frac{\pi}{L}\right)^{\Delta}\Psi_{-}(1)
+\mu_{+}\left(\frac{\pi}{L}\right)^{\Delta}\Psi_{+}(-1)\quad,
\end{eqnarray*}
where $L_{0}$ is the spectrum of the boundary conformal field theory
with central charge, $c$. The spectrum of $H(L)$ can be calculated
at least in two different ways: using perturbative (BCPT) and variational
methods (BTCSA). 

In the variational method we use, as input, the eigenvectors, $|n\rangle$,
of the unperturbed (boundary conformal) Hamiltonian. For practical
reasons we consider the eigenvectors having energy less then a given
value, $E_{cut}$, and perform the calculation numerically. (Technically
this means diagonalizing the truncated Hamiltonian).

Standard perturbation theory gives rise to the following perturbative
series for the energy level labeled with its unperturbed UV limiting
vector $|n\rangle$,
\begin{equation}
E_{n}(L)=\frac{\pi}{L}\left(E_{|n\rangle}-\frac{c}{24}
+\sum_{k=1}^{\infty}c_{k}^{n}(\mu_{bulk},\mu_{\pm},\Delta)
\left(\frac{\pi}{L}\right)^{k(\Delta-1)}\right)\quad \,,
\label{eq:En}
\end{equation}
where $E_{|n\rangle}$ denotes the conformal energy on the UHP. For the
ground state the first few terms have the form
\begin{equation}
E_{0}(L)=\frac{\pi}{L}\left(E_{|0\rangle}-\frac{c}{24}
+c_{1}^{0}(\mu_{\pm},\Delta)\left(\frac{\pi}{L}\right)^{(\Delta-1)}
+c_{2}^{0}(\mu_{bulk},\mu_{\pm},\Delta)\left(\frac{\pi}{L}\right)^{2(\Delta-1)}
+\dots\right)\quad,\label{eq:E0}
\end{equation}
where 
\begin{equation}
c_{1}^{0}(\mu_{\pm},\Delta)=\langle0|\left(\mu_{-}\Psi_{-}(1)
+\mu_{+}\Psi_{+}(-1)\right)|0\rangle\label{eq:c1}
\end{equation}
and 
\begin{eqnarray}
c_{2}^{0}(\mu_{bulk},\mu_{\pm},\Delta) & = & 
\mu_{bulk}\int_{0}^{\pi}\langle0|\Phi(e^{i\theta},e^{-i\theta})
|0\rangle d\theta\nonumber \\
 &  & +\sum_{i,j=\{\pm\}}\mu_{i}\mu_{j}\sum_{n\in\mathcal{H}}
\frac{\langle0|\Psi(i1)|n\rangle\langle n|\Psi(j1)|0\rangle}
{E_{|0\rangle}-E_{|n\rangle}}\quad.\label{eq:c2}
\end{eqnarray}
The large volume behavior of the ground state energy is 
\[
E_{0}(L)=\epsilon_{bulk}L+\epsilon_{boundary}^{-}
+\epsilon_{boundary}^{+}+O(e^{-mL})\quad\textrm{as \quad}L\to\infty\quad.\]
The ground state energy in the NLIE description, however, is normalized
differently as $E_{0}^{NLIE}(L)=O(e^{-mL})$ when $L\to\infty$. The
correspondence between the two schemes is 
\begin{equation}
E_{0}(L)=E_{0}^{NLIE}(L)+\epsilon_{bulk}L
+\epsilon_{boundary}^{-}+\epsilon_{boundary}^{+}\quad.
\label{eq:corresp} 
\end{equation}

\subsection{Lagrangian approach}\label{subsec:lagrangian}

The evaluation of the second order term in the Hamiltonian perturbation
theory (\ref{eq:c2}) is cumbersome, since we have to sum up the various
matrix elements. We can avoid this calculation by doing Lagrangian
perturbation theory instead. We compactify the strip in the time-like
direction on a circle of radius $R$ and consider the large $R$ limit
of the cylinder partition function: 
\[
Z(L,R)=Tr(e^{-RH(L)})=e^{-RE_{0}(L)}
+\dots\quad\textrm{for \quad}R\to\infty\quad.\]
Using the functional integral representation for the partition function
with the action, $S=\int L(t)dt=S_{BCFT}+S_{pert}$,
\begin{eqnarray*}
Z(L,R) & = & \int d[\Phi,\Psi_{\pm}]e^{-S}=
\int d[\Phi,\Psi_{\pm}]e^{-S_{BCFT}}e^{-S_{pert}}\\
 &=& Z_{BCFT}\frac{\int d[\Phi_{\ },\Psi_{\pm}]
 e^{-S_{BCFT}}e^{-S_{pert}}}{Z_{BCFT}}
 =Z_{BCFT}\left\langle e^{-S_{pert}}\right\rangle \\
 &=& Z_{BCFT}\left\langle \sum_{n=0}^{\infty}
 \frac{(-S_{pert})^{n}}{n!}\right\rangle \quad.
\end{eqnarray*}
We can obtain the first few perturbative corrections to the ground
state energy as
\begin{eqnarray*}
-\frac{1}{R}\lim_{R\to\infty}\log(Z(L,R)) & = & 
E_{0}(L)=E_{0}^{BCFT}(L)+\sum_{i=\pm}\mu_{i}\langle0|\Psi_{i}(0)|0\rangle\\
 &  & \hspace{-3.5cm}+\mu_{bulk}\int_{0}^{L}\langle0|\Phi(x,0)|0\rangle dx
 -\frac{1}{2}\sum_{i,j=\{\pm\}}\mu_{i}\mu_{j}
 \int_{-\infty}^{\infty}\langle0|\Psi_{i}(0)\Psi_{j}(t)|0\rangle dt+\dots\quad,
\end{eqnarray*}
where the correlators are the connected BCFT correlators. By transforming
the various expressions onto the upper half plane we obtain:\begin{eqnarray}
E_{0}(L) & = & E_{0}^{BCFT}(L)+\sum_{i}\mu_{i}\left(\frac{\pi}{L}\right)^{\Delta}
\langle0|\Psi_{i}(0)|0\rangle \label{eq:c2bound}\\
 &  & \hspace{-2.0cm}+\left(\frac{\pi}{L}\right)^{2\Delta-1}
 \left[\mu_{bulk}\int_{0}^{\pi}\langle0|\Phi(e^{i\theta},e^{-i\theta})
 |0\rangle d\theta-\frac{1}{2}\sum_{i.j}\mu_{i}\mu_{j}
 \int_{0}^{\infty}duu^{\Delta-1}\langle0|\Psi_{i}(i1)\Psi_{j}(ju)
 |0\rangle\right]-\dots \nonumber \end{eqnarray}
where $z=u+iv=re^{i\theta}$. Comparing the result with equations
(\ref{eq:E0},\ref{eq:c1},\ref{eq:c2}) we can establish the correspondence
with the Hamiltonian perturbation theory. Clearly the second order
term in (\ref{eq:c2}) is summed up. One can compare this term directly
by inserting the resolution of the identity $1=\sum_{n}|n\rangle\langle n|$
and using the conformal transformation property of the fields. 

In any BCFT, using the $sl_{2}$ invariance of the vacuum, $|0\rangle$,
the bulk one point function on the UHP can be put to the form
\begin{equation}
\langle0|\Phi(e^{i\theta},e^{-i\theta})|0\rangle=
\frac{c_{bulk}}{\sin(\theta)^{2\Delta}}\quad,\label{eq:cbulk}
\end{equation}
while the boundary two point function can be brought to the form
\begin{equation}
\langle0|\Psi_{i}(i1)\Psi_{j}((ju)|0\rangle
=\frac{c_{ij}}{|1-(i1)\cdot(ju)|^{2\Delta}} \,,
\label{eq:sl2inv}
\end{equation}
where the radial ordering is taken into account. The relevant
integrals can be written in terms of the beta function $B(x,y)$,
both for the bulk and for $i=j$ and for $i\neq j$ as 
\[
\int_{0}^{1}du\, u^{x-1}(1-u)^{y-1}=\int_{0}^{\infty}du\, 
u^{x-1}(1+u)^{-x-y}=B(x,y)=\frac{\Gamma(x)\Gamma(y)}{\Gamma(x+y)}\quad,\]
 The first integral converges only for $y>0$ thus $0<\Delta<\frac{1}{2}$
is needed. Collecting all terms, the $c_{2}^{0}$ coefficient is 
\begin{eqnarray}
c_{2}^{0}(\mu_{bulk},\mu_{\pm},\Delta) & = & \mu_{bulk}c_{bulk}
\frac{\Gamma(\frac{1}{2}-\Delta)\Gamma(\frac{1}{2})}
{\Gamma(1-\Delta)}\nonumber \\
 &  & \hspace{-1.5cm} - (\mu_{+}^{2}c_{++}+\mu_{-}^{2}c_{--})
 \frac{\Gamma(\Delta)\Gamma(1-2\Delta)}{\Gamma(1-\Delta)}
 -\mu_{-}\mu_{+}c_{+-}\frac{\Gamma^{2}(\Delta)}{\Gamma(2\Delta)}
 \quad,\label{eq:g0}
\end{eqnarray}
where only the coefficients $c_{bulk},\, c_{ij}$ are model dependent.

\subsection{Boundary sine-Gordon theory}\label{subsec:bsg}

In the sine-Gordon theory the UV limiting BCFT is described by
(\ref{lagrangian}),
the bulk perturbation is given by 
\[
\Phi(x,t)=\frac{1}{2}(V_{\beta}(x,t)+V_{-\beta}(x,t))\quad;
\quad V_{\beta}(z,\bar{z})=n(z,\bar{z}):e^{i\beta\varphi(z,\bar{z})}:\quad,\]
 while the boundary by 
 \[
\Psi_{\pm}(t)=\frac{1}{2}\left(e^{-i\frac{\beta\varphi_{0}^{\pm}}{2}}
U_{\frac{\beta}{2}}(t)+e^{i\frac{\beta\varphi_{0}^{\pm}}{2}}
U_{-\frac{\beta}{2}}(t)\right)\quad;\quad U_{\frac{\beta}{2}}(u)=:
e^{i\frac{\beta}{2}\varphi(u,u)}:\quad,\]
and $\Delta=\frac{\beta^{2}}{8\pi}$, see \cite{BPT2} for the details.
Since the bulk, $V_{\beta}(z,\bar{z})$, and boundary, $U_{\frac{\beta}{2}}(u)$,
vertex operators change the eigenvalue of $\pi_{0}$ by $\beta$ and
$\frac{\beta}{2}$, respectively, only even $c$ coefficients are
nonzero in the expansion (\ref{eq:En}). Moreover, the vacuum expectation
value of $\Phi$ is also zero (\ref{eq:cbulk}), thus the leading
perturbative contribution comes from the boundary two point function
part of (\ref{eq:c2bound}). In the boundary sine-Gordon theory with
nonzero $\kappa:=\kappa_{+}-\kappa_{-}$ the \emph{vacuum is not}
$sl_{2}$\emph{-invariant} and thus (\ref{eq:sl2inv}) has to be modified.
In general for a theory with a non $sl_{2}$-invariant vacuum one has
to compute the four point functions, instead of the two point function,
and extract the relevant matrix element from them. Alternatively, in
our case, one can also use the mode expansion of the field
(\ref{expansion})
together with the commutation relations (\ref{etcr}) to obtain: \[
\langle0|U_{\pm\frac{\beta}{2}}(i1)U_{\mp\frac{\beta}{2}}((ju)|0\rangle
=\frac{|u|^{\mp\frac{4\kappa\Delta}{\beta}}}{|1-(i1)\cdot(ju)|^{2\Delta}}\quad.\]
 This modifies (\ref{eq:g0}) and gives the leading corrections: 
 \begin{eqnarray}
c_{2}^{0}(\mu_{\pm},\Delta,\kappa) & = & -\frac{1}{2}
\left[(\mu_{+}^{2}+\mu_{-}^{2})\frac{\Gamma(1-2\Delta)}{2}
\left(\frac{\Gamma(\Delta+\frac{4\kappa\Delta}{\beta})}
{\Gamma(1+\frac{4\kappa\Delta}{\beta}-\Delta)}
+\frac{\Gamma(\Delta-\frac{4\kappa\Delta}{\beta})}
{\Gamma(1-\frac{4\kappa\Delta}{\beta}-\Delta)}\right)\right.\non \\
 &  & \left.+\mu_{-}\mu_{+}\cos\frac{\beta}{2}
 (\varphi_{0}^{+}-\varphi_{0}^{-})\frac{\Gamma(\Delta+\frac{4\kappa\Delta}{\beta})
 \Gamma(\Delta-\frac{4\kappa\Delta}{\beta})}{\Gamma(2\Delta)}\right]\quad.
\label{eq:finalresult} \end{eqnarray}
Although the derivation of this formula assumes that $0 <
\Delta < 1/2$, the final result is analytic in $\Delta$ (with possible
poles).  Therefore it has an analytic continuation for $\Delta > 1/2$,
which, since the NLIE is also analytic in $\Delta$, must coincide with
the NLIE result.  This is confirmed by experience with the NLIE for
bulk sine-Gordon and bulk supersymmetric sine-Gordon models.
Using the UV-IR relation (\ref{alyosha1}), (\ref{alyosha2}) and the 
mass-gap formula (cf. \cite{AZ2})
\begin{equation}
\mu_{bulk}=m^{2-2\Delta}\frac{2\Gamma(\Delta)}{\pi\Gamma(1-\Delta)}
\left(\frac{\sqrt{\pi}\Gamma\left(\frac{1}{2(1-\Delta)}\right)}
{2\Gamma\left(\frac{\Delta}{2(1-\Delta)}\right)}\right)^{2-2\Delta}\quad,
\label{eq:massgap}
\end{equation}
where $m$ is the soliton mass, $c_{2}^{0}(\mu_{\pm},\Delta,\kappa)$
can be rewritten in terms of the IR parameters.

\eject


\end{document}